\documentclass{article}[12pt]
\usepackage{amsfonts,inputenc,amssymb,amsmath,graphicx,geometry,xcolor,cite}
\usepackage{authblk, comment}
\usepackage{jheppub}
\usepackage{mwe}
\usepackage{float}
\usepackage{hyperref}
\usepackage{lipsum}
\usepackage{comment}
\newcommand{\mycomment}[1]{}
\title{Quantum Complexity of Nonlocal Field Theories}

\author[a]{Mohan sai Balusu,}
\author[b]{Gaurav Katoch,}
\author[c]{Sanhita Parihar}
\author[c]{and Shubho R. Roy}

\affiliation[a]{School of Physics, Indian Institute of Science Education and Research, 
695551 Thiruvananthapuram, India}
\affiliation[b]{Department of Physics, Indian Institute of Technology Indore, 453552 Indore, Madhya Pradesh, India}
\affiliation[c]{Department of Physics, Indian Institute of Technology, Hyderabad, Kandi, Sangareddy, Telengana 502285, India}

\emailAdd{balusumohansai19@alumni.iisertvm.ac.in}
\emailAdd{gauravk@iiti.ac.in}
\emailAdd{sanhita.hepth@gmail.com }
\emailAdd{roy.shubho@gmail.com}

\date{}

\abstract{Entanglement entropy for nonlocal field theories displays a universal ``volume law" scaling \cite{Barbon:2008ut, Karczmarek:2013xxa, Shiba:2013jja, Pang:2014tpa} as opposed to the ``area law" scaling for local field theories. The aim of this work is to determine whether complexity displays any such an universal scaling laws. The field theories considered here are obtained by deforming $\mathcal{N}=4$ SYM theory by higher dimension operators introducing nonlocality, namely a dipole deformation and noncommutativity (NCSYM) by turning on world volume Kalb-Ramond $B$ field. The dual gravity backgrounds have a running dilaton, in addition to the $B$-field background, which alter AdS asymptotics. Our results capture nonlocality in the hyperscaling behavior for complexity. We also compute the subregion complexity which display phase transitions in the nonlocal field theories with the transition point being the same as that for the phase transition of entanglement entropy \cite{Karczmarek:2013xxa}. These new results dovetail nicely with our findings from our previous works \cite{Chakraborty:2020fpt, Katoch:2022hdf, Bhattacharyya:2022ren} on other lower dimensional nonlocal field theories such as little string theories (LSTs) and warped conformal field theories (WCFTs).}

\begin{document}

\maketitle

\section{Introduction and Summary}
The gauge-gravity correspondence (AdS/CFT duality, BFSS (M)atrix model/M-theory)  \cite{Maldacena:1997re, Gubser:1998bc, Witten:1998qj, Aharony:1999ti, Itzhaki:1998dd, Banks:1996vh} has revolutionized our understanding of strongly coupled gauge theories and field theories in general. Strongly coupled regimes of field theories, apart from a few lower dimensional exceptional cases, are beyond the reach of coupling constant perturbation theory and other commonly used analytical tools. However, thanks to the gauge-gravity duality, they are now regularly being investigated by working with the weakly coupled gravitational dual system, often being perturbative (sugra) fields in weakly curved spacetime (bulk). The construction of the bulk can often be done in a ``\emph{bottom up}" fashion without appealing to details of the underlying string theory/M-theory compactifications or truncations). Thus, to obtain results or physical observables in strongly coupled field theories one effectively aims to solve (in most cases numerically) a much easier classical gravity-matter system, \emph{i.e.}\ Einstein field equations coupled to classical matter. This so called ``\emph{holographic approach}" of solving strongly coupled phenomena in fields theories has led to the pervasive use of GR/SUGRA tools in research areas as diverse as condensed matter physics/many-body physics \cite{Sachdev:2010ch, McGreevy:2009xe, Hartnoll:2009sz} to strongly coupled QCD/hadronic physics \cite{Erlich:2005qh, DaRold:2005mxj, Karch:2006pv}. It would be a gross oversimplification to state that the sole impact of gauge/gravity duality has been to provide a computational toolshed for strongly coupled regimes of field theory using Riemannian geometry. Far more impactful developments from gauge/gravity duality have been made while pondering how a completely nongravitational field theory encodes various phenomena on the gravity side, to wit, how the radial or holographic dimension emerge from field theory RG, how does the connectivity of the semiclassical bulk spatial dimensions arise, how does bulk causality arise in the semiclassical approximation, what codes the presence of event horizons in the bulk, and how does the dual field theory state inform us about formation of gravitational singularities in the semiclassical bulk, which dual field theory degrees of freedom contain knowledge of a specific bulk subregion and and so on. In fact, working on such issues led to the recognition of the significance of various concepts from the field of quantum information (and computation) (QIC) which are able to capture characteristics of quantum entanglement structure of states of quantum (field) theories that are not captured by traditional observables such as correlation functions of local operators, or even Wilson loop/t'-Hooft loop or surface operators. To name a few such QIC tools: Information geometry and information metrics, Entanglement or von-Neumann entropy \cite{Ryu:2006bv, Hubeny:2007xt} and associated generalization to Renyi \cite{Dong:2016fnf} Entropy, Mutual Information, Tensor networks and MERA \cite{Swingle:2009bg}, Fisher Information, Computational Complexity, Quantum error correction, and Fidelity susceptibility. The influx of concepts and computational tools from quantum information theory has turned out to be a game-changer, leading to unexpected insights in semiclassical gravity which has resolved some versions of the black hole information paradox \cite{Penington:2019npb, Almheiri:2019hni} and has led to very convincing proposals of reconstructing regions hidden behind black hole horizons \cite{Leutheusser:2021frk}. Since the advent of holography, one has been able to combine wisdom from diverse and often complementary techniques such as holography, integrability, supersymmetry (whenever applicable/available), lattice simulations, effective theories, and last but not the least good old perturbation theory, the landscape of local quantum field theories has been investigated rather thoroughly.  However, perhaps \emph{nonlocal} quantum field theories are still not as widely studied. A variety of nonlocal (noncommutative, nonisotropic or boost non-invariant) field theories arise in string theory as well as high energy physics in general. Many of these arise as low energy effective (UV incomplete) theories (e.g. \cite{Namsrai:1986md}) and some as UV complete theories (e.g.\cite{Sen:2015uaa}), and contrary to the folklore, such theories can turn out to be finite \cite{Efimov:1969fd} (even superrenormalizable) as well as unitary. Holography is turning out to be as productive in uncovering the strong coupling dynamics of nonlocal quantum field theories. Such exercise has an added benefit - this allows us to explore (quantum) gravity in spacetimes without AdS asymptotics, e.g. little string theories (LST) allows us to explore gravity in asymptotically flat spaces (with a dilaton). Thus nonlocal field theories might be a crucial role in proving nonperturbative descriptions of quantum gravity in more general spacetimes, not just asymptotically AdS spacetimes - a great motivation to study such theories.\\

The ultimate insight on holography from the quantum information theoretic approach is that the spacetime geometry is a code which represents the entanglement structure of the quantum state of the underlying dual nongravitational lower dimensional theory \cite{VanRaamsdonk:2009ar, VanRaamsdonk:2010pw}. The Ryu-Takayanagi (RT) prescription for entanglement entropy of the dual field theory state \cite{Ryu:2006bv, Hubeny:2007xt} was one of the first few evidences to shed light on this connection, while another one was Maldacena's proposal \cite{Maldacena:2001kr} of the eternal Schwarzschild-AdS (SAdS) geometry as the geometric representation of the \emph{thermofield-double} entangled state of two CFTs. Since then a mountain of evidence has accumulated relating a plethora of CFT observables which are metrics/diagnostics of quantum entanglement of the CFT state, to various geometrical/topological properties of the quasiclassical aAdS bulk (refer to \cite{VanRaamsdonk:2016exw} for a comprehensive review). Nevertheless, entanglement entropy or other entanglement measure such as tensor networks/MERA, error-correcting codes do not appear to describe the geometry of bulk regions which are concealed by event horizons (black hole interiors). The interior geometry in the bulk is highly time-dependent, while the dual field theory is time-independent (in equilibrium). For example, consider the case of the Einstein-Rosen Bridge (ER bridge) or nontraversable wormholes in the interior of an SAdS black hole. The entanglement entropy of the boundary CFT saturates in a short time upon attaining thermal equilibrium but in the bulk the ER bridge keeps on growing linearly with Schwarzschild time (apparently eternally) even though the dual field theory attains thermalization. To account for the eternal linear-in-time growth of the ER bridge, Susskind  \cite{Susskind:2014rva} has resorted to another key idea from quantum information theory and added it to lexicon of holography, to wit, the \emph{computational complexity} or better yet the \emph{quantum complexity} of the dual field theory state. Quantum Complexity is a characteristic of the states in the Hilbert space of a quantum mechanical or field theoretic system, which quantifies the level of ``difficulty" in preparing that state (dubbed the ``target state") from a given initial state (dubbed the ``reference state") using a specified set of operators (called ``gates")\footnote{Truth be told, there is no guarantee that one will be able to reach the target state exactly, and so one also has to introduce a ``tolerance" parameter, quantifying the proximity to the target state one can get or gets to while using the specified set of gates the Hilbert space}. While such a definition is perfectly implementable for a quantum mechanical system with discrete degrees of freedom, like quantum circuits in information theory, it turns out to be enormously difficult to have a well-defined UV-finite quantum complexity in the continuum limit, i.e. those described by a (quantum) field theory. This is not uncommon, while taking the continuum limit, many quantities which are well-defined in the discretum case cease to be well-defined in the continuum. As a result, a mathematically well-defined (UV-finite) and unanimously accepted definition of quantum complexity is still lacking for continuum field theories. We recall the geometric scheme of Nielsen et. al.\cite{2005quant.ph..2070N, 2006Sci...311.1133N}, whereby a quantum circuit complexity for a field theory is prescribed to be the minimum number of unitary gates in the space of unitary operators described by general Finsler metrics. As a result, the complexity of a target state with respect to a given reference state, can now be prescribed to be length of the minimal geodesic in a Finsler metric-space (with appropriate cost functions and penalty factors as free parameters). The distance or length functional now acts like action integral i.e. a variational problem. The cost functions are mandated to obey physical restrictions/conditions e.g. continuity and differentiability as well as those necessary to define a distance function such as positivity, symmetry and the triangle inequality. Despite the obvious geometric appeal and other virtues such as UV-completeness in this Nielsen approach, there is still enormous freedom in the choosing cost functions or penalty factors determining the Finsler metric and hence in the quantum complexity of the state itself. Also not every pair of points can be joined by geodesics in general in a given manifold, so there is no guarantee that Nielsen complexity can be defined at all for a given pair of reference and target states. A considerable number of attempts have been made to define complexity in the continuum limit (see e.g \cite{Jefferson:2017sdb, Chapman:2017rqy, Khan:2018rzm, Yang:2018nda, Molina-Vilaplana:2018sfn, Hackl:2018ptj, Bhattacharyya:2018wym, Guo:2018kzl, Bhattacharyya:2018bbv, Yang:2018tpo, Camargo:2019isp, Balasubramanian:2019wgd, Bhattacharyya:2019kvj, Erdmenger:2020sup, Bueno:2019ajd, Chen:2020nlj, Flory:2020eot, Flory:2020dja} for an incomplete list). Still, no universal, unanimous definition of quantum complexity exists at the moment in the continuum limit, and neither is a complete classification of the possible universality classes is available at the moment. Actually, in the continuum limit, quantum complexity, by definition is a (UV) divergent object as it is only defined to within a tolerance (say $\epsilon$) with respect to the target state. It is counterproductive to demand further precision in reproducing the target state, as it entails inclusion of more number/quantity of gates, which depends inversely tolerance, and thus is a divergent quantity. Traditionally UV divergent quantities, or quantities which explicitly depend on the UV-cutoff are considered to be of no physical interest in continuum QFT, and understandably so as their value can be adjusted to anything by just redefining the UV cutoff. However, it is the characteristic UV-cutoff dependence, that is the indisputable defining feature of quantum complexity in QFT e.g. the exponents which specify the degree of the divergence (much akin to critical exponents in statistical systems). \\
	
Early on, two distinct prescriptions were put forward by Susskind et. al. with distinct motivations, with regard to which geometrical construction represents the bulk dual to the  dual boundary (CFT) state complexity. The very first one, posits that the CFT state complexity to be represented by the volume of a codimension-one (spacelike) maximal volume slice in the bulk aAdS which is anchored at the exact same boundary spatial slice (time) on which the CFT (boundary) quantum state is given \cite{Susskind:2014rva} up to some universal constant and a suitably selected length scale characterizing the bulk (say the AdS radius or the event horizon radius). Naturally this was dubbed the complexity-volume ($CV$) proposal or prescription. The underlying motivation comes from the fact that the tensor networks representing the many body CFT state resemble discretizations of negatively curved spatial slices (hyperbolic) i.e. those bulk AdS. And in terms of tensor networks complexity is given by the size of the tensor network, which in turn is a discretization of the volume of the bulk spatial slice. The second conjecture or prescription \cite{Brown:2015bva, Brown:2015lvg} stipulates that the CFT state complexity is proportional to the on-shell bulk (sugra) action integral confined to the Wheeler-deWitt (WdW) patch of the boundary spatial slice on which the dual CFT state is given. The WdW patch of a specific boundary spatial slice is the union of all possible spacelike surfaces in the bulk which are anchored to the very same boundary spatial slice. This conjecture was dubbed the complexity-action ($CA$) proposal. Both these candidate duals share the common traits of linear growth in time, extensivity (scales linearly with boundary volume) and demonstrate the switchback effect. Since the asymptotically AdS or even flat bulk is a noncompact space, both these candidate bulk geometric duals of CFT state complexity are UV divergent, and one has to impose a IR/radial cutoff in the bulk, a standard practice in holography. In the CV proposal the length scale characteristic of the geometry, which was introduced for dimensional consistency also introduced an ambiguity. For the CA proposal, there are also couple of issues. The WdW patch has boundaries, some of whom are codimension-one null hypersurfaces, and further their \emph{with} edges/junctions. Such null boundaries and their junctions necessitated  the inclusion of suitable boundary terms in the WdW patch action as demonstrated in \cite{Lehner:2016vdi}. Recently, however, an enormous generalization of the original two proposals has been obtained whereby a whole class of different geometric constructions, based on both codimension-one and codimenension-zero subregions, all of which display linear growth, boundary volume law divergence, and the switchback effect \cite{Couch:2016exn, Belin:2021bga, Belin:2022xmt}. For reviews on quantum complexity in the context of QFT and Quantum Gravity (Holography) see \cite{Chapman:2021jbh, Baiguera:2025dkc}.\\
	
In past work \cite{Chakraborty:2020fpt, Katoch:2022hdf, Bhattacharyya:2022ren} we investigated the quantum complexity of \emph{Little String theory} (LST) (obtained from the decoupled limit of the (nonlocal) theory describing a stack of $k$ NS$5$-branes (in the limit of large $k$, ($k\gg1$)) wrapping $T^4 \times S^1$) and the somewhat related warped CFTs, both being nonlocal field theories in $1+1$ dimensions. The reason of the nonlocality of the LST i.e. NS$5$ brane worldvolume theory is that it decouples from the bulk  at \emph{finite} string length $l_s=\sqrt{\alpha'}$, i.e. the LST on the NS$5$ branes, still retains stringy nonlocality! As has been widely reported, LST provides an excellent theoretical laboratory to study nonlocal (nongravitational) field theories, as it is intermediate between string theory (which is nonlocal theory containing massless gravitons) and a local QFT. The dual SUGRA geometry is extracted by the near horizon geometry of the NS$5$-brane stack - a flat metric accompanied by a linear dilaton profile, $\mathbb{R}^{1,1}\times \mathbb{R}_\phi$ (the dilaton blowing up at spatial infinity). This duality has been studied thoroughly over the years \cite{Aharony:1998ub,Kutasov:2001uf}. Further, wrapping $p$ F$1$ strings ($p\gg 1$) on a $S^1$ along the NS$5$-brane worldvolume, the near-horizon geometry of the F-strings turns into AdS$_3$. Thus the dual sugra background flows from AdS$_3$ in the IR (corresponding to the near-horizon geometry of the F strings) to flat spacetime linear dilaton in the UV (which corresponds to the near horizon geometry of just the NS$5$ branes). As a result, the dual nongravitational boundary field theory interpolates between  a (local) CFT$_2$ in the IR to (nonlocal) LST in the UV. This duality is concrete example of holography in non-AdS background although it is related by an RG flow to $AdS_3$. Naturally being an instance of a non-AdS holography, has gained the attention in many works, e.g.see \cite{Asrat:2017tzd,Chakraborty:2018kpr,Chakraborty:2018aji,Chakraborty:2020xyz,Chakraborty:2020udr,Chakraborty:2020yka} for a few examples. In our recent work we probed this theory using holographic complexity as a probe i.e. we computed the volume and action complexity, both at zero and finite temperature. The complexity expressions has signatures of the intrinsic nonlocality owing to the stringy excitations on the UV divergence structure. In particular, we obtained quadratic and logarithmic divergences, which cannot arise in local field theory in $1$ space dimension (instead of a linear divergence consistent with extensivity for a local field theory) when the UV cutoff is smaller than the so-called Hagedorn length/energy scale, set by the $T\overline{T}, J\overline{T}, T\overline{J}$ couplings. The UV divergence structure of the quantum complexity also contained the signature of Lorentz (boost) violation, although the Lorentz violation does not introduce any novel UV divergence, just like nonlocality it contributes to the quadratic and logarithmic divergences. We further worked out the subregion volume complexity which displayed a phase transition, which is identified with the Hagedorn transition, with the transition point being the same as that for the phase transition of entanglement entropy. We also studied finite temperature corrections to the complexity, however we did not discover newer or more exotic UV-divergences up to second order. \\

The purpose of this paper is to extend the scope our work to more general setting, i.e. using holography to study more examples of (boundary) nonlocal field theories which naturally arise in the string theory with nonvanishing dilaton and $B$-field turned on. These are all theories obtained in the decoupling limit of a system of branes with a nonzero $B$-field and dilaton turned on. So these theories are only AdS in the deep IR, while in the UV they depart from AdS asymptotics (due to the dilaton). The non-zero $B$ field on the worldvolume renders the dual field theories nonlocal (noncommutative). We study two such examples exhaustively, the first being the dipole deformed $\mathcal{N}=4$ SYM theory \cite{Bergman:2000cw,Chakravarty:2000qd,Dasgupta:2000ry,Bergman:2001rw,Dasgupta:2001zu}, which can be obtained from an irrelevant deformation of $\mathcal{N}=4$ SYM theory by a (nonlocal) vector operator of dimension five \cite{Dasgupta:2000ry} in the low-energy limit.
This has a 10 dimensional gravity dual via a string theory realization \cite{Bergman:2001rw}. In the UV, the dynamics of this theory is expected to be in a phase that is dominated by strong (attractive) dipole interactions and one of our aims to use holographic complexity to probe that regime. The second example we study in detail is the Noncommutative Super Yang-Mills theory (NCSYM) obtained by putting the Super-Yang Mills theory in a noncommutative spacetime \(\mathbb{R}^2_\theta \times \mathbb{R}^{1,1}\). In a string theory setting this is naturally realized by taking the decoupling limit of the worldvolume theory  of D3-branes in the presence of NS--NS \(B\)-field as was shown in \cite{Hashimoto:1999ut, Maldacena:1999mh, Alishahiha:1999ci}. In this case as well, the dual SUGRA (bulk) background has a dilaton and a B-field turned on, which modify the AdS asymptotics in the UV. The impact of noncommutativity on the dual field theory is profound. The noncommutativity also induces anisotropy as is evident from the target space. This theory displays characteristic trait of noncommutative field theories in general, namely UV/IR mixing -- high energy (UV) physics becomes entangled with long-distance (IR) physics, due to non-local interactions. Using holography, in particular, quantum information theoretic measures, such as holographic entanglement entropy \cite{Karczmarek:2013xxa}, reveal modified scaling laws and phase structures for the NCSYM. We aim to build of these work by studying holographic complexity to uncover the various phases of this theory. A major motivation of our work, i.e. to study such exotic (nonlocal, noncommutative) field theories is because they shed light on holography in non-AdS spaces. In general, it appears that to study holography in spacetimes with asymptotically non-AdS boundaries, e.g. null boundaries, one needs to study such exotic field theories \cite{Ferko:2025elh, Dei:2025ryd, Chakraborty:2025rfm}.\\

The plan of the paper is as follows. In Sec. \ref{DDSYM} we consider the dipole deformed $\mathcal{N}=4$ SYM (DDSYM). We first provide a self-contained but extremely brief recap of the salient features of the theory and present the dual SUGRA background containing a nontrivial dilaton, a $B$-field and a Ramond-Ramond $4$-form potential. This is particularly simple example of
noncommutative theory as it is free from UV/IR mixing. We investigate the complexity characteristics of the dipole deformed SYM via three holographic probes, namely full volume complexity, the subregion volume complexity as a function of the subregion size (length along the dipole direction), and finally the action complexity using the full 10d SUGRA action. In the following section, Sec. \ref{NCSYM1}, we tackle the case of the more complicated noncommutative SYM theory, where the base space consists of the noncommutative space. First we consider the noncommutative base space to contain the two-dimensional noncommutative plane $\mathbb{R}^2_\theta$. In this case, UV-IR mixing is inevitable. Also, as a result, one necessarily has to deal with effects of anisotropy. Later on in Sec. \ref{NCSYM-iso} we consider the fully isotropic SYM, for which a holographic bulk (SUGRA) dual was provided in \cite{Maldacena:1999mh, Lin:2000ps}. For both the anisotropic and isotropic case, we work out the holographic full volume complexity as well as the action complexity using the $10$d and $5$d actions. Studying both cases is essential to disentangle the effect of anisotropy from noncommutativity. Interestingly, for the action complexity, it appears that one has to incorporate a topological term in the onshell action, first prescribed by \cite{Kurlyand:2022vzv} (also see \cite{Apolo:2025wcl} for a recent work involving this term). For the convenience of the reader, we summarize our findings here:\\
\begin{itemize}
    \item DDSYM case: We find that the total volume complexity is \emph{independent} of the dipole-deformation parameter and hence naturally matches with the pure AdS result. We conjecture that this is a dimensional accident (the fact that the dipole deformation is anisotropic and chooses a preferred direction) which is supported by the holographic computation, where the factors $f(u)$ containing the deformation parameter $a$ gets canceled out from the overall bulk volume (complexity) functional even before extremizing. In contrast, the subregion complexity, which contains a tunable scale, namely the subregion size (length), \emph{does} depend on the deformation parameter, especially in strongly nonlocal UV regime (small subregion size) but reduces to pure AdS (local) volume law near IR (large subregion). Our numerical study reveals a phase transition which is same as the one recorded by holographic entanglement entropy at the same critical subregion size (length). Finally, the action complexity analysis leads to a constraint on t'Hooft coupling/dipole-deformation parameter $a$ in order for complexity to remain positive, in strongly local IR regime. Also it coincides with the volume complexity when the t'Hooft coupling is one. 
    \item Anisotropic NCSYM case: Analogous to the dipole deformed theory, here the total volume complexity remains independent of the noncommutativity parameter $a$, which we attribute to be a dimensional accident (it is obvious from the holographic perspective as the metric factors noncommutativity parameter cancel out in the volume complexity functional even before extremizing it). The subregion volume complexity, on the other hand, exhibits strong $a$-dependence with distinct scaling behaviors due to the introduction of a tunable scale, namely the subregion size along the direction of noncommutativity. In the near-AdS regime (IR), the complexity admits perturbative corrections to pure AdS complexity, with the subleading corrections being quartic in $a$ (reducing to local field theory volume law in the $a\rightarrow0$ limit). In the far-AdS regime, the complexity displays nonlocal hyperscaling proportional to the fourth power of subregion length and sixth power of the noncommutativity parameter, manifesting UV/IR mixing characteristic of noncommutative theories. A phase transition emerges at critical length scale set by the inverse square of $a$, coinciding precisely with the entanglement entropy phase transition point. Finally, the action complexity, just like the dipole deformed theory, leads to constraint on the noncommutative coupling parameter for complexity to remain positive. Although, action complexity agrees with the volume law (cubic) scaling  of the volume complexity and the subregion complexity, but it involves logarithmic dependence upon the nonlocality parameter, the feature missing from the volume complexities.   
    
    \item Isotropic NCSYM case: Interestingly in this case the full volume complexity \emph{does} depend on the isotropic noncommutativity parameter $a$. This is in contrast to both the dipole deformed and anisotropic NCSYM case discussed above where the deformation singled out one particular direction confirming our conjecture that the independence of complexity from deformation parameters in the previous two cases were dimensional accidents. In the weakly noncommutative regime, the complexity at leading order is identical to pure AdS but receives perturbative corrections, which are quartic in powers of $a$. In the strongly noncommutative regime, the complexity displays a nonlocal hyperscaling dependence on $a^{28/3}$. On the other hand, the  action complexity also predicts the hyperscaling dependence in $a$ and the UV scale,  it  disagrees with the scaling exponents of both, the coupling parameter and the UV scale of the volume complexity. Moreover, the action complexity is manifestly an arbitrarily large  negative number hinting to the fact that perhaps the isotropic NCSYM is UV incomplete \textit{viz.} it is UV unstable. This is also in contrast to the behavior of the complexity of anisotropic case as discussed earlier.     
\end{itemize}

In Sec. \ref{DO} we discuss our results in the context of past work, the scope and limitations of our approach, and provide an outlook for future work. In the Appendix \ref{EE_dipole}, we revisit the holographic entanglement entropy of the dipole deformed SYM theory to facilitate comparison with the subregion volume complexity, especially to determine the critical subregion length as a function of the deformation parameter and the fact that the critical subregion size turns out to the same for both observables - be it entanglement entropy or subregion volume complexity. In Appendix \ref{app: num_error} we furnish some details regarding the numerical limitations we encountered while performing the computations for the isotropic NCSYM case using Mathematica. \\

{\bf{Note}}: We set the AdS radius to unity ($R_{AdS}=1$) throughout in what follows.

\section{Dipole Deformed SYM theory} \label{DDSYM}
As the first example of a nonlocal field theory, we consider a simple noncommutative field theory, namely the dipole deformed $\mathcal{N}=4$ SYM theory (DDSYM) \cite{Bergman:2000cw,Chakravarty:2000qd,Dasgupta:2000ry,Bergman:2001rw,Dasgupta:2001zu}. The dipole deformation of a given commutative (undeformed) field theory can be constructed by replacing the ordinary product of the two local fields $\phi_{i}$ and $\phi_{j}$ by a \emph{star product}, to wit,
\begin{eqnarray}\label{eq: starprod}
    \phi_{i}(\vec{x})\star \phi_{j}(\vec{x})= e^{\left(L_{i}^{\mu}\frac{\partial}{\partial x^{\mu}}-L_{j}^{\nu}\frac{\partial}{\partial x^{\nu}}\right)} \phi_{i}(x)\phi_{j}(x) = \phi_{i}\left(\vec{x}-\frac{\vec{L}_{j}}{2}\right) \phi_{j}\left(\vec{x}+\frac{\vec{L}_{i}}{2}\right)
\end{eqnarray}
where, $\vec{L}_{i}$ is a vector assigned to each field $\phi_{i}$, known as dipole vector.
As a result of this deformation, the conformal symmetries as well supersymmetries are broken and one expects the properties of the theory at low energies to be more on lines of usual gauge theories e.g. a IR confined phase.
In momentum space this replacement by star product is equivalent to a simple phase shift,
\begin{equation}
    \text{Tr}\left(\phi_{1} (p_1)\star\phi_{2}(p_{2})\star...\star \phi_{n}(p_{n})\right)=e^{\sum_{1\le i<j\le n}L_{j}p_{i}}\text{Tr}\left(\phi_{1}(p_{1})\phi_{2}(p_{2})...\phi_{n}(p_{n})\right)
\end{equation}
where sum of all the dipoles vanishes\footnote{This leads to the fact that all planar contribution in the deformed theory is same as the ordinary theory.} $\sum_{i}L_{i}=0=\sum_{i}p_{i}$, due to cyclicity of trace.
In general, when multiple fields are involved in the product there can be ordering ambiguity, but as we are interested in gauge theories there is a natural ordering imposed by the gauge group. The requirement of associativity over the star products, constrain the dipole length associated with the star product of two fields to be sum of the dipole length of individual fields\footnote{A general choice of the dipole vector which ensure this is $\vec{L}_{i}=Q_{i}\vec{L}$, where $Q_{i}$ are the conserved charges and $\vec{L}$ is a constant vector.},
\begin{eqnarray}
    (\phi_{i}\star\phi_{j})\star\phi_{k}&=&\phi_{i}\left(\vec{x}-\frac{\vec{L}_{j}}{2}\right)\phi_{j}\left(\vec{x}+\frac{\vec{L}_{i}}{2}\right)\star\phi_{k}(\vec{x})\nonumber\\
    &=&\phi_{i}\left(\vec{x}-\frac{\vec{L}_{j}}{2}-\frac{\vec{L}_{k}}{2}\right)\phi_{j}\left(\vec{x}+\frac{\vec{L}_{i}}{2}-\frac{\vec{L}_{k}}{2}\right)\phi_{k}\left(\vec{x}+\frac{L_{i}+L_{j}}{2}\right)\nonumber\\
    &=&\phi_{i}\left(\vec{x}-\frac{(\vec{L}_{j}+\vec{L}_{k})}{2}\right)\left(\phi_{j}\star\phi_{k}\right)\left(\vec{x}+\frac{\vec{L}_{i}}{2}\right)\nonumber\\
    &=&\phi_{i}\star(\phi_{j}\star\phi_{k})
\end{eqnarray}
and the CPT invariance in the dipole theory identifies the dipole length associated with the hermitian conjugate of field $\phi^{\dagger}$ to $-\vec{L}$.

From here on, for simplicity, we take $\vec{L}_{i}=\vec{L}_{j}=L\,\hat{x}$, for a fixed length scale $L$, so that respective theory is nonlocal only in the $x$-direction. 
It has been shown in \cite{Dasgupta:2000ry} that in low energy limit the dipole deformed $\mathcal{N}=4$ SYM theory is obtained from an irrelevant deformation of $\mathcal{N}=4$ SYM theory by a (nonlocal) vector operator of dimension five.
Later in \cite{Bergman:2001rw} a 10 dimensional gravity dual of dipole theory is obtained by a string theory realization\footnote{In type IIB string solution with probe D3 brane, the dual geometry has been constructed by applying T-duality along an compactified brane direction followed by a twist in $S^{5}$ directions and finally another T-duality along the same direction. In \cite{Maldacena:2008wh} a particular example of this geometry has been identified with the non-relativistic background introduced in \cite{Son:2008ye}, and its thermodynamic properties has been studied.}
\begin{eqnarray}\label{eq: dipolebulk}
     ds^2 &=&  u^2 \left ( -dt^2 + f(u) dx^2 + dy^2
 +dz^2 \right ) +
\frac {du^2}{u^2} +d\tilde{\Omega}_{5}^{2}~,\\ \nonumber
d\tilde{\Omega}_{5}^{2}&=&\frac{|d\alpha|^2+|d\beta|^2}{1+|\alpha|^2+|\beta|^2}
-\frac{|\bar\alpha d\alpha + \bar \beta d\beta|^2}
{(1+|\alpha|^2+|\beta|^2)^2}
+ f(u)d\psi^{2}
 \nonumber\\
 d\psi&=&\left (d \tilde{\gamma} +
\frac{\text{Im}(\bar\alpha d\alpha + \bar \beta d\beta)}
{1+|\alpha|^2+|\beta|^2}
\right )\nonumber\\
e^{2\phi} &=&  g_s^2 ~f(u)=e^{2\phi_{0}}f(u) ~,\\ 
B_{x\psi} &=& - \frac{1-f(u)} {\tilde L} = -\frac{ a u^2 f(u)}{\tilde{L}}~,\\ \nonumber
C_{txyz}&=&u^{4}\nonumber\\
f(u) &=& \frac{1}{1 + (a u)^2}  ~.\nonumber   
\end{eqnarray}
where the bulk background parameter $a=s \tilde L$ is related to the length scale of nonlocality (noncommutativity) of the boundary dual field theory $L$ as $L=2\pi \tilde{L}$, with $s$ being the t'Hooft coupling $s=4\pi g_{YM}^{2}N$. 
This dual geometry is a deformation (warping) of $AdS_{5}\times S^{5}$ along one AdS direction $x$ and one $S^{5}$ direction by a factor of $f(u)$. The deformation on the $S^{5}$ is such that the radius of $S^{1}$ fibration over $C\mathbb{P}^{2}$ is modified to $f(u)$. It is evident from the metric, dilaton and B-field expressions that in the commutative regime ($a\rightarrow 0$) we recover the usual $AdS_{5}\times S^{5}$ background with vanishing dilaton and B-field. The generic features of this geometry has been studied in detail in \cite{Bergman:2001rw}, and they have shown the boundary metric is degenerate and this degeneracy sources the nonlocality along the $x$-direction in the boundary theory. Also this theory is free from any UV/IR mixing, which are present in noncommutative Super-Yang-Mills theory(NCSYM), this makes it a simpler example of non-commutative theory. While the study of correlation function in this background remains an open problem, a study of Wilson loops of the dipole theories has been carried out using the gravity dual in \cite{Alishahiha:2002ex, Huang:2007ux,Huang:2007dg}, and it has shown presence of an dipole-dipole interaction term in the interquark potential, in addition to the usual gauge theory coulomb potential. This dipole-dipole interaction produces an attractive force between gluons within a glueball and also lead to an additional confining force between quarks in baryons, which shows an indication of new baryon phase due to this dipole-dipole attraction \cite{Huang:2008uh}. Various aspects of the dipole deformed gauge theory has been studied \cite{Chiou:2003sh,Fischler:2013gsa,Karczmarek:2013xxa} over the years using holographic tools and several interesting results has been found. In the study of entanglement entropy in dipole deformed theory \cite{Karczmarek:2013xxa} it has been found that entanglement entropy follows a volume law in the region smaller than the non-commutative length scale rather than an area law. This leaves the study of behavior of complexity in the dipole deformed theory as an attractive outstanding problem, which might shed more light on the phase structure of the dipole deformed gauge theory. In the remainder of this section, we investigate the complexity characteristics of the dipole deformed gauge theory via three holographic probes, namely full volume complexity, the subregion volume complexity as a function of the subregion size (in particular the dependence of the subregion complexity on the length or extension along the dipole deformation) and finally the action complexity (using the full 10d SUGRA action). Overall, we observe that while in the near commutative/ near local regime (small $a$) the complexity does reproduce local volume law scaling(s), in the strongly nonlocal regime (large $a$), complexity departs from volume law scaling with the scaling exponent being \emph{independent} of the dipole deformation parameter $a$! We comment on our findings at the end of the section.
\subsection{Volume complexity (CV)}
In this section we study the nature of complexity in the dipole deformed theory by following the ``holographic complexity =Volume" proposal\cite{Susskind:2014rva}, which states that the complexity of the boundary theory is given by the volume of an maximal volume spacelike hypersurface.
\begin{eqnarray}
    \mathcal{C}_{V}=\frac{V_{\Sigma}}{G_{10}\ell},\quad V_{\Sigma}=\int d^{d-1}x\sqrt{\gamma}
\end{eqnarray}
where,$\gamma$ is the pullback metric on the hypersurface, $\ell$ is a characteristic length scale, and $G_{10}$ is ten dimensional Newton's constant.  
Since here we are working in string frame and have a non-trivial dilaton background we use a modified volume functional \cite{Klebanov:2007ws},
\begin{eqnarray}
    \mathcal{C}=\frac{V_{\Sigma}}{G_{10}R},\quad V_{\Sigma}=\int d^{d-1}x~ e^{-2(\phi-\phi_{0})}\sqrt{\gamma}
\end{eqnarray}
where, $\phi$ is the dilaton field, $e^{\phi_{0}}=g_{s}$ is the string coupling constant, and we have taken the characteristic length scale to be same as AdS radius.
To proceed we find the induced metric on a spacelike constant time hypersurface ($t'(u)=0$),
\begin{eqnarray}
    ds^2 &=&  \frac{  \left(1-u^4 t'(u)^2\right)}{u^2}du^2+  u^2 f(u)dx^2+ u^2dy^2+ u^2 dz^2+d\tilde{\Omega}_{5}^{2}
\end{eqnarray}
Then the volume of the hypersurface is given by,
\begin{eqnarray}
    V_{t}&=&\int du\, dx\, dy\, dz\, e^{-2(\phi-\phi_{0})}~\text{Vol}(\tilde{\Omega}_{5})\sqrt{h}\nonumber\\
     V_{t}&=&\int du\, dx\, dy\, dz\, \pi ^3 u^2 \sqrt{1-u^4 t'(u)^2}
\end{eqnarray}
where, the volume of the deformed sphere $\text{Vol}(\tilde{\Omega}_{5})=\pi^{3}\sqrt{f(u)}$ and frame change introduce an additional factor $f(u)^{-1}$. For the extremal surface $t'(u)=0$ the volume reduced to,
\begin{eqnarray}
     V_{t}&=&\frac{\pi ^3 W^3 u_{b}^{3}}{3}
\end{eqnarray}
where $u_{b}$ is the boundary UV-cutoff and the $W$ is the boundary IR cut off (the limits of integration for $-W/2\leq x,y,z\leq W/2$). The simplified maximal volume determines the complexity,
\begin{equation}
    \mathcal{C}_{V}=\frac{ u_{b}^3 W^3}{3 G_{5}}
\end{equation}
where we have used the relation between 10 dimensional Newton constant and 5 dimensional Newton constant $G_{10}=\text{Vol}(S_{5})G_{5}=\pi^{3}G_{5}$.

Following AdS/CFT dictionary the five dimensional Newton constant (in units of AdS radius) is given in terms of field theory data, $G_{5}=\frac{\pi}{2N^{2}}$
\begin{eqnarray}
    \mathcal{C}_{V}=\frac{2N^{2}}{3\pi}\left(\frac{W}{\epsilon}\right)^{3}=\frac{8c}{3\pi}\left(\frac{W}{\epsilon}\right)^{3}
\end{eqnarray}
where, $c$ is the central charge $c=\frac{N^{2}}{4}$ in large $N$ limit and we have introduced the ``lattice spacing", $\epsilon$, as an UV regulator in the boundary via the bulk IR cutoff $u_{b}=\frac{1}{\epsilon}$.

From the resultant expression of volume complexity of the dipole deformed theory we notice that it is not only independent of the deformation parameter $a$, but it is exactly same as the volume complexity of pure AdS \cite{Reynolds:2016rvl}. This happens due the exact cancellation of combined deformation factors from the deformed $x$-direction and deformed sphere with the frame factor, which is very specific to the spacetime dimensions we have considered the dipole deformed theory. 
\subsection{Subregion volume complexity}
In this section we will study the subregion complexity of the dipole deformed theory, namely the infinitely long strip: s $\frac{-W}{2}<y,z<\frac{W}{2}$ and $\frac{-l}{2}<x<\frac{l}{2}$ with $W\rightarrow\infty$. 
The subregion complexity will be given by the maximal volume of co-dimension one spacelike surface bounded by the co-dimension two Ryu-Takanayagi(RT) surface, homologous to the subregion \cite{Alishahiha:2015rta}.
The metric on co-dimension two surface,
\begin{eqnarray}
    ds^{2}=\frac{ \left(u^4 f(u)+u'(x)^2\right)}{u^2}dx^2+u^2 dy^2 + u^2 dz^2 
\end{eqnarray}
and the area of the RT surface $u(x)$, is given by
\begin{eqnarray}
    \mathcal{A}&=&\int e^{-2(\phi-\phi_{0})}\text{Vol}(\tilde{\Omega}_{5})\sqrt{h_{s}}\nonumber\\
    \mathcal{A}&=&\pi ^3  W^{2}\int_{\frac{-l}{2}}^{\frac{l}{2}}dx\, \left(\frac{u \sqrt{u^4 f(u)+u'(x)^2}}{\sqrt{f(u)}}\right) \nonumber\\
    \mathcal{A}&=& \pi ^3 W^{2}\int_{\frac{-l}{2}}^{\frac{l}{2}}dx\, \mathcal{L}(u,u')
\end{eqnarray}
From this point we can proceed in two ways: we can either directly perform the definite integral involved in the area functional it will give us area of the RT surface as a function of subregion length $\mathcal{A}(l)$, or a further simplification can be made by observing that here the functional $\mathcal{L}(u,u')$ does not explicitly depends on the co-ordinate $x$ that means there is a conserved quantity associated with this invariance. This conserved quantity let' say $H(u(x),u'(x))$ can be used to replace $u'(x)$ in terms of the extremum value of $u(x)$ which we denote by $u_{*}$ \emph{i.e.} $u=u_{*}$ when $u'(x)=0$. This replacement after a variable change from the boundary coordinate $x$ to bulk coordinate $u$ simplifies the integral involved in the area functional and leaves us with both area functional and subregion length in terms of extremum value of u, $\mathcal{A}(u_{*})$ and $l(u_{*})$. We can still study area of the RT surface as a function of subregion length $\mathcal{A}(l(u_{*})$ by varying $u_{*}$, below we proceed in this way

\begin{figure}
    \centering
    \includegraphics[width=0.4\linewidth]{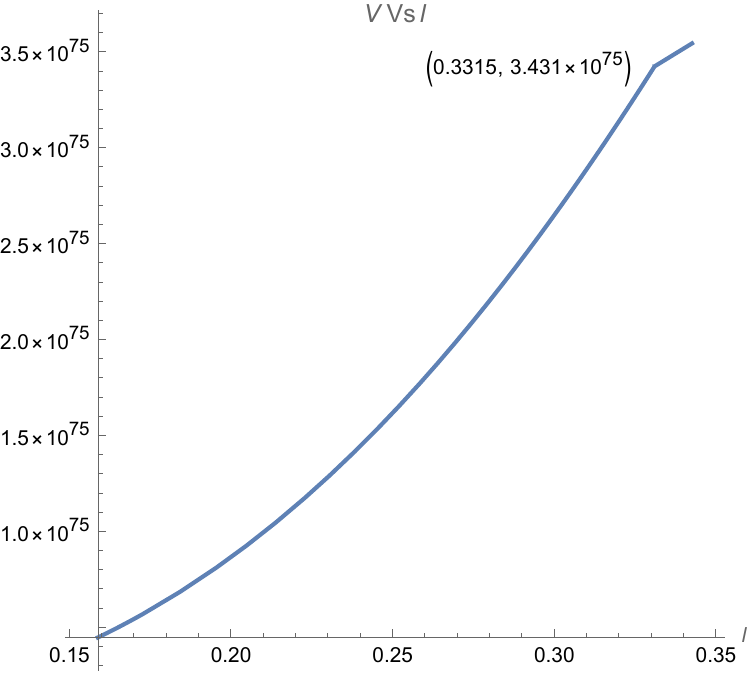}\hspace{0.1cm}    \includegraphics[width=0.4\linewidth]{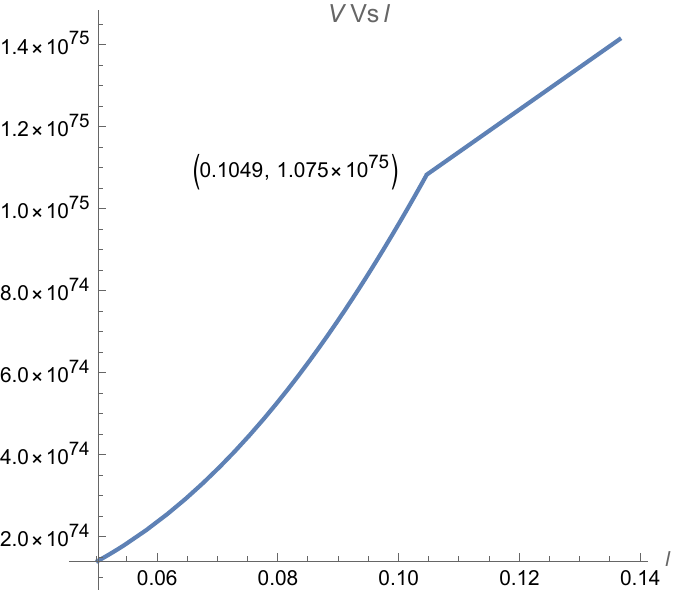}    \includegraphics[width=0.4\linewidth]{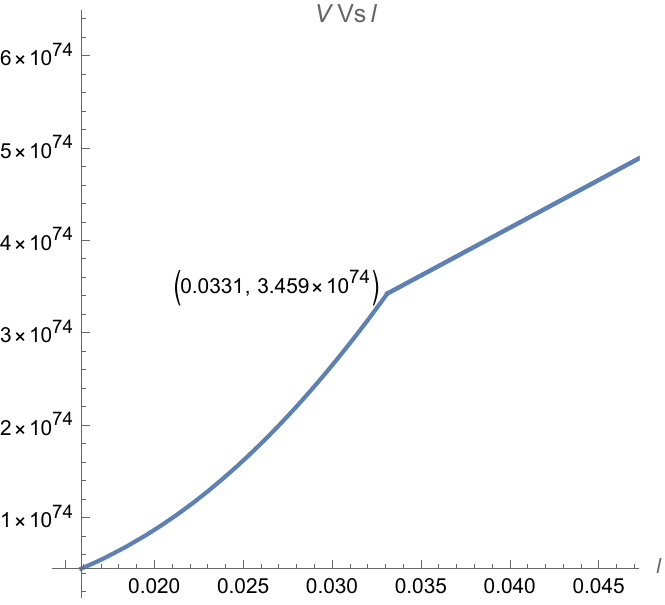}    \includegraphics[width=0.4\linewidth]{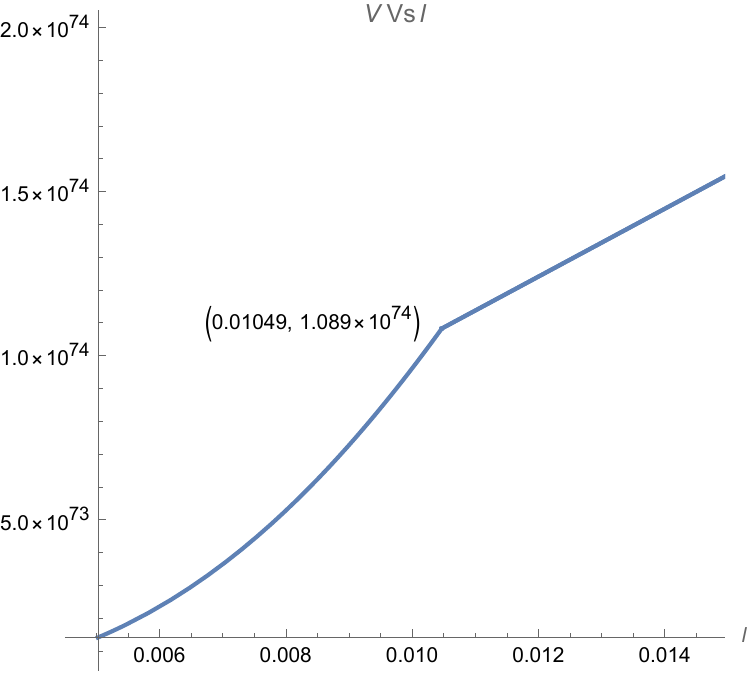}    \includegraphics[width=0.4\linewidth]{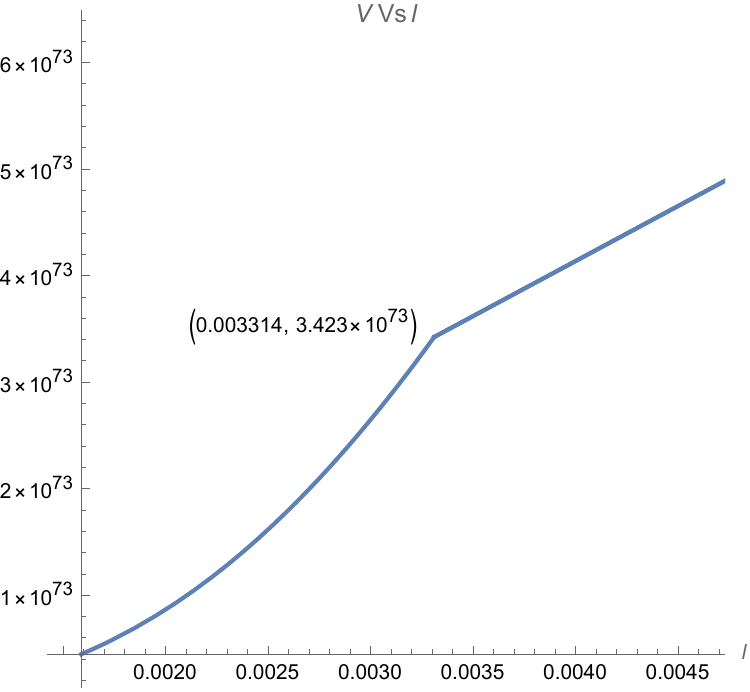}\hspace{0.3cm}    \includegraphics[width=0.4\linewidth]{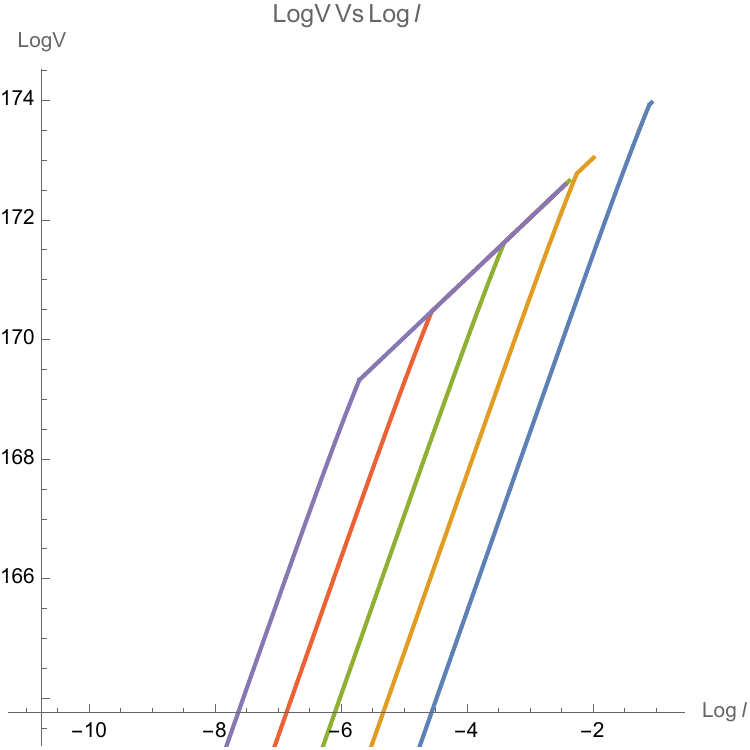}
    \caption{Different numeric plots of $C_{V}$ Vs $l$ for five different values of $a=10^{-0.5},10^{-1.0},10^{-1.5},10^{-2.0},10^{-2.5}$ and $W=10^{15}$ and in the last we have shown the numeric plots of $\ln(C_{V})$ Vs $\ln(l)$.  For all the plots we have considered AdS radius $R=1$ and $W=10^{15}$, and pointed out the co-ordinates for the transition point.}
    \label{fig:Cvvsl_dipole}
\end{figure}

The conserved Hamiltonian is given by,
\begin{eqnarray}\label{eq : DT_Hamiltonian}
    H=u'\frac{\partial \mathcal{L}}{\partial u'}-\mathcal{L=}-\frac{u^5 \sqrt{f(u)}}{\sqrt{u^4 f(u)+u'(x)^2}}=-u_{*}^{3}
\end{eqnarray}
From \eqref{eq : DT_Hamiltonian} we can replace $u'(x)$ in terms of $u_{*}$, this simplifies the area of co-dimension two surface,
\begin{eqnarray}\label{eq: DT_area}
               \mathcal{A}&=&\pi ^3 W^2 \int_{u_{*}}^{u_{b}}du\, \frac{ u^4\sqrt{a^2 u^2+1}}{\sqrt{u^6-u_{*}^6}}
\end{eqnarray}
Also, the subregion length $l$ can be found in terms of $u_{*}$
\begin{eqnarray}\label{eq : DT_subregionlength}
    l=\int_{-\frac{l}{2}}^{\frac{l}{2}} dx=2\int_{u_{*}}^{u_{b}}du\, \frac{u_{*}^{3}\sqrt{a^2 u^2+1}}{u^2 \sqrt{u^6-u_{*}^6}}
\end{eqnarray}
The integral involved in \eqref{eq: DT_area} and \eqref{eq : DT_subregionlength} refrain us to proceed further analytically.A full numerical analysis of the area/entanglement entropy and a comparison of final results with \cite{Karczmarek:2013xxa} has been presented in appendix \ref{EE_dipole}. Here we focus on the subregion complexity, and consider the volume of co-dimension one surface bounded by RT surface. 
\begin{eqnarray}\label{eq : subregionComp_dipole}
    \mathcal{V}&=&\int_{u_{*}}^{u_{b}} du\, \int_{-x(u)}^{x(u)}dx\, \int_{-\frac{W}{2}}^{\frac{W}{2}}dy\,\int_{-\frac{W}{2}}^{\frac{W}{2}} dz\, e^{-2(\phi-\phi_{0})}\text{Vol}(\tilde{\Omega}_{5})\sqrt{h}\nonumber\\
    \mathcal{V}&=&2\pi ^3 W^{2}\int_{u_{*}}^{u_{b}} du\,  u^2 \int_{u_{*}}^{u} d\tilde{u}\, \frac{u_{*}^3 \sqrt{a^2 \tilde{u}^2+1}}{\tilde{u}^2 \sqrt{\tilde{u}^6-u_{*}^6}} 
\end{eqnarray}
We could not proceed analytically with the above integral as well, therefore we move forward with numerics, and analyze the parametric behavior of subregion complexity $\mathcal{V}(u_{*})$ with the finite subregion length $l(u_{*})$, shown in figure \ref{fig:Cvvsl_dipole}. 
From the numeric plots in \ref{fig:Cvvsl_dipole} it is clear that there are two phases of the system and both have different dependence on the subregion length $l$. The critical length at which this transition happen can be found from \eqref{eq : DT_subregionlength} in the limiting case  of far from AdS $(a u)>>1$, when the extremal surface lies near to the boundary,
\begin{eqnarray}
    L_{c}&=&2\int_{u_{*}}^{u_{b}} du\, \frac{u_{*}^{3}a }{u \sqrt{u^6-u_{*}^6}}\nonumber\\
    L_{c}&=&\frac{\pi a}{3}+\mathcal{O}\left(\frac{1}{u_{b}}\right)
\end{eqnarray}
As shown in the table \ref{tab:DT_critical length} this critical value of subregion length exactly matches the subregion length corresponding to the phase transition point in the numeric plots. We can compute subregion complexity analytically in certain limiting regions far from AdS $(au>>1)$ and near AdS $(au<<1)$.
\begin{table}[]
    \centering
    \begin{tabular}{|c|c|c|}
    \hline
    Deformation Parameter ($a$) & $L_{c}=\frac{a\pi}{3}$ &$L_{c}$ from Plot\\
    \hline
     $10^{-0.5}$&0.331153   &0.3315\\
     \hline
    $10^{-1.0}$& 0.10472  &0.1049\\
    \hline
    $10^{-1.5}$& 0.033115   &0.0331\\
    \hline
    $10^{-2.0}$&0.010472  &0.01049\\
    \hline
    $10^{-2.5}$& 0.003311  &0.00331\\ 
    \hline
    \end{tabular}
    \caption{ In this table we have shown the values of critical length for different values of deformation parameter $a$ (in units of AdS radius) and compared them with the critical subregion length from the $C_{V}$ Vs $l$ numeric plots}
    \label{tab:DT_critical length}
\end{table}
\subsubsection{Far from AdS \texorpdfstring{$(a u>>1)$}:}
In this limit the integrals involved in \eqref{eq : DT_subregionlength} and \eqref{eq : subregionComp_dipole} can be solved exactly. The subregion length is given by,
\begin{eqnarray}
    l=\frac{2}{3} a \cos ^{-1}\left(\frac{u_{*}^3}{u_{b}^3}\right);\quad u_{b}>u_{*}>0
\end{eqnarray}
and the sub-region complexity is,
\begin{align}
     \mathcal{C}_{s}=&\frac{2 a N^2 W^2}{9 \epsilon ^3 \pi} \Bigg(\pi-4 \sin ^{-1}\left(\frac{\sqrt{1-\sin \left(\frac{3 l}{2 a}\right)}}{\sqrt{2}}\right)\nonumber\\&-2 \cos \left(\frac{3 l}{2 a}\right) \left(\ln \left(\sin \left(\frac{3 l}{2 a}\right)+1\right)+\ln \left(\sec \left(\frac{3 l}{2 a}\right)\right)\right)\Bigg) 
\end{align}
where, $u_{b}=\frac{1}{\epsilon}$.

When the sub-region length is much less than the deformation parameter \emph{i.e.} $l<<a$ then,
\begin{eqnarray}
 \mathcal{C}_{s}=\frac{ a N^2 W^2 }{2 \pi \epsilon ^3} \left(\frac{l}{a}\right)^{3}+\mathcal{O} \left(\left(\frac{l}{a}\right)^{7}\right) \label{DDSYM cubic law}
\end{eqnarray}
Evidently this is a violation of the volume law behavior ($W^2l^3=V\,l^2$). The anisotropy of the dipole deformation contributes to this form of a scaling, i.e. volume times square of the extension along the dipole deformation. Instead of focusing on the transverse directions (those which are unaffected by the dipole deformation), we focus on the dependence of the subregion extension along the deformation $l$, then we get a cubic scaling law.
\subsubsection{Near AdS \texorpdfstring{$au<<1$}:}
In the near AdS region we can proceed by performing the integrals exactly and the respective sub-region length and complexity is given by,
\begin{eqnarray}
    l&=&\frac{2 \sqrt{\pi } \Gamma \left(\frac{2}{3}\right)}{u_* \Gamma \left(\frac{1}{6}\right)}-\frac{u_*^3 \, _2F_1\left(\frac{1}{2},\frac{2}{3};\frac{5}{3};\frac{u_*^6}{u_b^6}\right)}{2 u_b^4}\nonumber\\
        \mathcal{C}_{s}&=&\frac{ N^2 W^2}{3 \pi u_*} \left(\frac{u_*^4 \left(4 \, _2F_1\left(\frac{1}{6},\frac{1}{2};\frac{7}{6};\frac{u_*^6}{u_b^6}\right)-\, _2F_1\left(\frac{1}{2},\frac{2}{3};\frac{5}{3};\frac{u_*^6}{u_b^6}\right)\right)}{u_b}+\frac{\sqrt{\pi } \left(\Gamma \left(\frac{5}{3}\right) u_b^3+\frac{12 u_*^3 \Gamma \left(\frac{7}{6}\right)^2}{\Gamma \left(-\frac{1}{3}\right)}\right)}{\Gamma \left(\frac{7}{6}\right)}\right).
\end{eqnarray}
Large $u_{b}$ limit,
\begin{eqnarray}
    \mathcal{C}_{s}&=&\frac{ N^2 W^2}{3 \pi u_*} \left(\sqrt{\pi}\frac{ \Gamma \left(\frac{5}{3}\right) u_b^3}{ \Gamma \left(\frac{7}{6}\right)}+\sqrt{\pi}\frac{12 \Gamma \left(\frac{7}{6}\right)}{\Gamma \left(-\frac{1}{3}\right)}\right)+\mathcal{O}\left(\frac{1}{u_{b}}\right)\\
    l&=&\frac{2 \sqrt{\pi } \Gamma \left(\frac{2}{3}\right)}{u_* \Gamma \left(\frac{1}{6}\right)}
\end{eqnarray}
leads to
\begin{eqnarray}
    \mathcal{C}_{s}=\frac{ N^2}{3 \pi }\frac{l W^{2}}{\epsilon^{3}}+\frac{4 \pi ^{3/2} N^2 W^2}{3 \sqrt{3} l^2 \Gamma \left(-\frac{2}{3}\right) \Gamma \left(\frac{1}{6}\right)}+\mathcal{O}(\epsilon)
\end{eqnarray}
Expectedly this is a volume law ($\propto l\,W^2$) behavior (linear in dipole-deformed extension $l$).
\subsection{Action Complexity}
In this section we compute the complexity of the dipole theory, by following the action complexity conjecture, according to which the complexity of the boundary theory is given by the on-shell action of the dual bulk theory on the Wheeler-De-Witt (WDW) patch \cite{Brown:2015bva,Brown:2015lvg}.
\begin{equation}
    \mathcal{C}_{\mathcal{A}}=\frac{\mathcal{I}_{WDW}}{\pi \hbar}
\end{equation}
The WDW patch is an union of all spacelike hypersurfaces anchored at a boundary time (let's say $T$) bounded by the null hypersurfaces.
\begin{eqnarray}
     ds^{2}&=&0\nonumber\\
     dt&=&\pm\frac{du}{u^{2}}\nonumber\\
     t(u)&=&2\int_{u}^{u_{b}}\frac{du}{u^{2}}=2\left(\frac{1}{u}-\frac{1}{u_{b}}\right)
\end{eqnarray}
It can also be regarded as the domain of dependence of the time slice which we consider in CV conjecture.
In general the action on WDW patch consists of contributions from the bulk action, boundary action on the null boundaries of WDW patch, joint terms from the intersection of two null boundaries\footnote{We regularize the action by shifting the edge of the WDW patch to the regulator surface and hence we only have a single null joint at this time slice.} and a counter term added to the action to make sure that action is invariant under the reparameterization of null generators\cite{Lehner:2016vdi}
\begin{equation}
    \mathcal{I}_{WDW}= \mathcal{I}_{bulk}+\mathcal{I}_{bdy}+\mathcal{I}_{joint}+\mathcal{I}_{LMPS}
\end{equation}
\textbf{Bulk on-shell action:}
Now, as the known 10D bulk dual\cite{Bergman:2001rw} of the dipole theory is a solution of type  IIB supergravity theory we consider type IIB SUGRA action,
\begin{align}
    S_{IIB}&=S_{NS}+S_{RR}+S_{CS}\\
    \text{where,}\hspace{1cm }S_{NS}&=\frac{1}{2\tilde{\kappa}_{10}^2}\int d^{10}X \sqrt{-g}e^{-2\Phi}\left(\mathcal{R}+4\partial_{\mu}\Phi\partial^{\mu}\Phi-\frac{1}{2}|H_{(3)}|^2\right)\\
    S_{RR}&=-\frac{1}{4\tilde{\kappa}_{10}^2}\int d^{10}X\sqrt{-g}\left(|F_{(1)}|^2+|\tilde{F}_{(3)}|^2+\frac{1}{2}|\tilde{F}_{(5)}|^2\right)\\
    S_{CS}&=-\frac{1}{4\tilde{\kappa}_{10}^2}\int C_{(4)}\wedge H_{(3)} \wedge F_{(3)}
\end{align}
where,
\begin{equation}
    \int d^{10}X \sqrt{-g}|F_{(p)}|^{2}=\frac{1}{p!}\int d^{10}X \sqrt{-g}\, g_{\mu_{1}\nu_{1}}...g_{\mu_{p}\nu_{p}}F^{\mu_{1}\mu_{2}...\mu_{p}}F^{\nu_{1}\nu_{2}...\nu_{p}}
\end{equation}
\begin{eqnarray}
    2\tilde{\kappa}^{2}_{10}&=&16\pi G_{10}e^{-2\phi_{0}}=16\pi^{4}G_{5}e^{-2\phi_{0}}=2\kappa_{10}e^{-2\phi_{0}},\quad F_{(p)}=dC_{(p-1)},\quad H_{(3)}=dB_{(2)}\nonumber\\
    \tilde{F}_{(3)}&=&F_{(3)}-C_{(0)}H_{(3)},\quad \tilde{F}_{(5)}=F_{(5)}-\frac{1}{2}C_{(2)}\wedge H_{(3)}+\frac{1}{2}B_{(2)}\wedge F_{(3)}
\end{eqnarray}
with the on-shell solution \eqref{eq: dipolebulk}.

The bulk dual geometry of dipole theory only has one non-vanishing Ramond-Ramond five-form field strength $F_{5}$, therefore on-shell,
\begin{eqnarray}
    S_{RR}&=&-\frac{1}{4\tilde{\kappa}_{10}^2}\int d^{10}X\sqrt{-g}\left(\frac{1}{2}|F_{(5)}|^2\right)\nonumber\\
    S_{CS}&=&0
\end{eqnarray}
But, as the RR five form field is constrained to be self dual \cite{Sen:2015nph}, $\star F_{5}=F_{5}$ the on-shell kinetic action of RR-field also gives vanishing contribution, and only non-vanishing contribution comes from $S_{NS}$, which consist of,
\begin{itemize}
        \item Einstein Hilbert action:\footnote{Here we have presented the final complexity contributions redefined by a prefactor $\mathcal{C}_{\mathcal{A}}=(2\pi\kappa_{10} \hbar) \mathcal{C}_\mathcal{A}$ }
        \begin{eqnarray}
        S_{EH}&=&\frac{1}{2\tilde{\kappa}_{10}^2}\int d^{10}X \sqrt{-g}e^{-2\Phi}\mathcal{R}\nonumber\\
        \mathcal{C}_{\mathcal{A}}^{EH}&=&W^{3}\int_{0}^{u_{b}} du\int_{-t(u)}^{t(u)}dt~u^3\frac{2 a^2 u^2 \left(8 a^2 u^2+15\right)}{\left(a^2 u^2+1\right)^2}\\
       &=& -\frac{10 W^3 \tan ^{-1}\left(a u_b\right)}{a^3}-\frac{2 W^3 u_b}{a^2}+\frac{12 W^3 \ln \left(a^2 u_b^2+1\right)}{a^4 u_b}+\frac{8}{3} W^3 u_b^3
    \end{eqnarray}
        \item Kinetic action of dilaton action:
        \begin{eqnarray}
             S_{\phi}&=& \frac{4}{2\tilde{\kappa}_{10}^2}\int d^{10}X \sqrt{-g}e^{-2\Phi}g^{MN}\partial_{M}\phi\partial_{N}\phi\nonumber\\
             &=&        \frac{4}{2\tilde{\kappa}_{10}^2}\int d^{10}X \sqrt{-g}e^{-2\Phi}g^{uu}\partial_{u}\phi\partial_{u}\phi\nonumber\\
        \mathcal{C}_{\mathcal{A}}^{\phi}&=&4W^{3}\int_{0}^{u_{b}} du\int_{-t(u)}^{t(u)}dt~u^3\frac{a^4 u^4}{(a^2 u^2+1)^{2}}\nonumber\\
        &=&\frac{20 W^3 \tan ^{-1}\left(a u_b\right)}{a^3}-\frac{8 W^3 u_b}{a^2}-\frac{12 W^3 \ln \left(a^2 u_b^2+1\right)}{a^4 u_b}+\frac{2}{3} W^3 u_b^3
        \end{eqnarray}
        \item Kinetic action of B-field:
        \begin{eqnarray}
S_{H}&=& - \frac{1}{4\tilde{\kappa}_{10}^2}\int d^{10}X \sqrt{-g}e^{-2\Phi}\frac{1}{3!} g^{\mu_{1}\nu_{1}}g^{\mu_{2}\nu_{2}}g^{\mu_{3}\nu_{3}}H_{\mu_{1}\mu_{2}\mu_{3}}H_{\nu_{1}\nu_{2}\nu_{3}}\nonumber\\
       \mathcal{C}_{\mathcal{A}}^{H}&=&- \frac{W^{3}}{2}\int_{0}^{u_{b}} du\int_{-t(u)}^{t(u)}dt~u^3\frac{24 a^4 u^2 \left(2 a^2 u^2+3\right)}{\tilde{L}^{2} \left(a^2  u^2+1\right)^2}\nonumber\\ 
       &=&-\frac{4 a^2 W^3 u_b^3}{\tilde{L}^2}-\frac{12 W^3 \tan ^{-1}\left(a u_b\right)}{a \tilde{L}^2}+\frac{12 W^3 u_b}{\tilde{L}^2}
        \end{eqnarray}

\end{itemize}
\textbf{Null boundary action contribution:} 
The general contribution of null boundary is given by,
\begin{eqnarray}
    S_{bdy}=-(\pm) \frac{1}{\tilde{\kappa}_{10}}\int d^{8}X\, ds\,\sqrt{-\tilde{\gamma}}\, \tilde{\kappa}
\end{eqnarray}
where, $(\pm)$ depends on the if volume of interest lies in future or past of the null boundary, $\gamma$ is the induced metric on the null hypersurface, $s$ parameterize the null generators, and $\kappa$ is given by,
\begin{equation}
    n^{\mu}\nabla_{\mu}n^{\nu}=\kappa n^{\nu},\quad n^{\mu}=\frac{\partial x^{\mu}}{\partial s}
\end{equation}
where, $n^{\mu}$ is null normal vector on the hypersurface, and $\kappa$ measures the failure of $s$ to be an aﬃne parameter.
This boundary contribution is valid in Einstein frame, while the bulk dual of dipole theory is known in string frame. Therefore we need to transform the quantities involved in string frame through a conformal transformation. As in 10D the bulk metric in string frame and Einstein frame are related as,
\begin{eqnarray}
   \tilde{g}_{\mu\nu}&=&e^{-\frac{\phi}{2}}g_{\mu\nu}=\Omega^{2}g_{\mu\nu} \nonumber\\
    \tilde{\kappa}&=&\kappa+2 n^{\mu}\nabla_{\mu}\ln \Omega \\
    \sqrt{\tilde{\gamma}}&=&\Omega^{8}\sqrt{\gamma}
\end{eqnarray}
As we anchor the WDW patch at some constant time $T$, the WDW patch has two regions when $t>T$ the upper half of WDW patch $\mathcal{N}^{+}$ and when $t<T$ the lower half of the WDW patch $\mathcal{N}^{-}$.

The total boundary contribution,
\begin{eqnarray}
    S_{bdy}&=&S_{\mathcal{N^{+}}}+S_{\mathcal{N^{-}}}
\end{eqnarray}
\begin{eqnarray}
    S_{\mathcal{N^{+}}}&=&\frac{1}{\tilde{\kappa}_{10}}\int d^{8}X\, ds_{+}\,\sqrt{-\tilde{\gamma}}\, \tilde{\kappa}_{+}\nonumber\\
    S_{\mathcal{N^{-}}}&=&-\frac{1}{\tilde{\kappa}_{10}}\int d^{8}X\, ds_{-}\,\sqrt{-\tilde{\gamma}}\, \tilde{\kappa}_{-}\nonumber\\
\end{eqnarray}
We have chosen the affine parameter $s_{\pm}=\mp u$, so the null normal vector is such that $n^{\mu}_{\pm}=\mp\frac{1}{R}\frac{\partial x^{\mu}}{\partial u}=\alpha_{\pm}(\frac{1}{u^{2}},\mp1,\vec{0})$ it gives a vanishing $\kappa=0$. Therefore only contribution to the boundary action comes due to the frame change,
\begin{eqnarray}
    \tilde{\kappa}_{\pm}=2 n^{\mu}_{\pm}\nabla_{\mu}\ln \Omega=\mp\frac{a^2 \alpha_{\pm} u}{2 \left(1+a^2 u^2\right)}
\end{eqnarray}
\begin{equation}
S_{bdy}=S_{\mathcal{N}^{+}}+S_{\mathcal{N}^{-}}=2S_{\mathcal{N}^{+}}
\end{equation}
taking $\alpha_{+}=\alpha_{-}=\alpha$
\begin{equation}
            \mathcal{C}_{\mathcal{A}}^{bdy}=\alpha\left(\frac{2 W^3 \tan ^{-1}\left(a u_b\right)}{a^3}-\frac{2 W^3 u_b}{a^2}+\frac{2}{3} W^3 u_b^3\right)
\end{equation}
\textbf{Joint contribution:}
The joint contribution coming from the intersection of null boundaries $\mathcal{N}^{+}$ and $\mathcal{N}^{-}$ at $u=u_{b}$, is given by,
\begin{eqnarray}
      S_{J_{1}}=-\frac{1}{\tilde{\kappa}_{10}^{2}}\int d^{8}X ~\Omega^{8}\sqrt{-\gamma} ~\ln\left(\frac{\vec{n_{+}}.\vec{n_{-}}}{2}\right)  
\end{eqnarray}

\begin{eqnarray}
    g_{\mu\nu}n_{+}^{\mu}n_{-}^{\nu}=-\frac{2\alpha_{+}\alpha_{-}}{u_{b}^{2}}
\end{eqnarray}
this gives 
\begin{eqnarray}
    \mathcal{C}^{joint}_{\mathcal{A}}&=&-2W^3 u_{b}^3\ln \left(\frac{\alpha_{+}\alpha_{-}}{ u_{b}^2}\right)\
\end{eqnarray}
This is exactly same as the joint term contribution in case of pure AdS \cite{Reynolds:2016rvl}. Again taking $\alpha_{+}=\alpha_{-}=\alpha$

\begin{equation}
    \mathcal{C}_{\mathcal{A}}^{joint}=4 W^3 u_b^3 \ln \left( u_b\right)-4W^3 u_{b}^3\ln(\alpha)
\end{equation}
 \textbf{LMPS counter term:}
 To retain the action on WDW patch to remain invariant under reparameterization of parameter $s$, we need to add an counter term known as LMPS counter term\cite{Lehner:2016vdi} given by,
\begin{equation}
    \Delta S=-(\pm)2\int d^{8}X ds\sqrt{\tilde{\gamma}}~ \Theta \ln{(l_{ct}~ \Theta)},\quad \Theta=\frac{1}{\sqrt{\gamma}}\frac{\partial\sqrt{\gamma}}{\partial s}
\end{equation}
where, the plus and minus sign is determined based on volume of interest lies in the future or past of the considered null boundary, and $l_{ct}$ is an arbitrary scale. 
\begin{eqnarray}
\Delta S_{\mathcal{N}^{+}}&=&-2\int d^{8}X ds^{+}\sqrt{\tilde{\gamma}}~ \Theta_{+} \ln{| \Theta_{+}|},\quad \Theta_{+}=\frac{1}{\sqrt{\tilde{\gamma}}}\frac{\partial\sqrt{\tilde{\gamma}}}{\partial s^{+}} \nonumber\\
    \Delta S_{\mathcal{N}^{-}}&=&2\int d^{8}X ds^{-}\sqrt{\tilde{\gamma}}~ \Theta_{-} \ln{| \Theta_{-}|},\quad\Theta_{-}=\frac{1}{\sqrt{\tilde{\gamma}}}\frac{\partial\sqrt{\tilde{\gamma}}}{\partial s^{-}}\nonumber    
\end{eqnarray}
As we have chosen the affine parameter, $ s^{\pm}=\mp u$
    \begin{equation}
      \Theta_{\pm}=\mp\frac{3}{R u}  
    \end{equation}
with this,
\begin{eqnarray}
        \mathcal{C}_{\mathcal{A}}^{LMPS}&=&\frac{4}{3} W^3 u_b^3+4 W^3 \ln (3) u_b^3-4 W^3 u_b^3 \ln \left( u_b\right)+4W^{3}u_{b}^{3}\ln(l_{ct})
\end{eqnarray}
\textbf{Topological counter term:}
The AdS limit of the bosonic part of the 10d type IIB action when evaluated on the vacuum $AdS_5\times S^5$ paradoxically vanish and is inconsistent  with the value of the would be 5d action obtained after it had been compactified over $S^5$. Similar disagreement had also been noted between the on-shell values of the reduced 3d action and the 10d action found in the case of $AdS_3 \times  S^3 \times \mathcal{T}^4$ background supported by
a 3-form flux. A natural way to resolve this problem has been suggested in \cite{Kurlyand:2022vzv} where, it is assumed that the 10d action is missing some
topological boundary term that restores the equivalence of its on-shell value with that of the 5d action. The additional boundary topological term is given by, 
\begin{eqnarray}
    S_{top}&=&\frac{1}{4\tilde{\kappa}_{10}^2}\int d^{10}X\sqrt{-g}\left(|F^{(e)}_{(5)}|^2\right)\nonumber\\
    &=&-\frac{1}{4\tilde{\kappa}_{10}^2}\int ~F_{(5)}^{(e)}\wedge\star F_{(5)}^{(e)}
\end{eqnarray}
As there is an non-zero five-form RR field supporting the dual bulk geometry of dipole theory \eqref{eq: dipolebulk},
\begin{eqnarray}
    F_{(5)}^{(e)}&=&dC_{4}\nonumber\\
    &=&4 u^{3}~dt\wedge du\wedge dx \wedge dy\wedge dz
\end{eqnarray}
this term adds a non-trivial contribution to the complexity 
\begin{eqnarray}
    \mathcal{C}_{\mathcal{A}}^{top}&=&-8 W^3\int_{0}^{u_{b}}du\int _{-t(u)}^{t(u)}dt~u^{3}\nonumber\\
    &=&-\frac{4}{3}u_{b}^3 W^3
\end{eqnarray}
which is the expected bulk contribution arising from pure AdS \cite{Reynolds:2016rvl}.\\

Summing over all contributions, the total action complexity turns out to be,

\begin{eqnarray}
    \mathcal{C}_{\mathcal{A}}&=&\frac{2 \alpha W^3 \tan ^{-1}\left(\frac{a}{\epsilon}\right)}{a^3}-\frac{12 s ^2 W^3 \tan ^{-1}\left(\frac{a}{\epsilon}\right)}{a^3}+\frac{10 W^3 \tan ^{-1}\left(\frac{a}{\epsilon}\right)}{a^3}-\frac{2 \alpha W^3 }{a^2\epsilon}+\frac{12 s ^2 W^3 }{a^2\epsilon}\nonumber\\&&-\frac{10 W^3 }{a^2\epsilon}+4 \left(\frac{W}{\epsilon}\right)^3 \log (\ell)+\frac{2}{3} \alpha \left(\frac{W}{\epsilon}\right)^3-4 \left(\frac{W}{\epsilon}\right)^3 \log (\alpha) -4 s ^2 \left(\frac{W}{\epsilon}\right)^3+\frac{10}{3} \left(\frac{W}{\epsilon}\right)^3\nonumber\\&&+4 \log (3)\left(\frac{W}{\epsilon}\right)^3
\end{eqnarray}
where in the final expression we have written the UV cutoff $u_{b}$ in terms of lattice spacing $u_{b}=\frac{1}{\epsilon}$. If we fix the normalization constant to be one $(\alpha=1)$ and the counter term scale to be AdS radius which has been taken to be one $(l_{ct}=R=1)$ the expression of complexity simplifies to,
\begin{align}
    \mathcal{C}_{\mathcal{A}}=&4 \left(\frac{W}{\epsilon}\right)^3\ln 3+\frac{12 W^3 \tan ^{-1}\left(\frac{a}{\epsilon}\right)}{a^{3}}\left(1-s^{2}\right)-\frac{12 W^3 }{a^{2}\epsilon}\left(1-s^{2}\right)+4 \left(\frac{W}{\epsilon}\right)^3\left(1-s^{2}\right) \label{DDSYMCA}
\end{align}
where, $s$ is t'Hooft coupling. This is a remarkable feature of DDSYM theory that the action complexity has an explicit dependence on the 't-Hooft coupling parameter. Regarding this, we would like to point out that the dipole deformation parameter is a function of the 't-Hooft coupling parameter, namely $a=s\,\tilde{L}$, and consequently appears in the classical sugra background (metric, running dilaton and NS-NS B-field) itself! This is why any observable in the sugra regime, be it sugra field correlators, geodesics, holographic wilson loops, holographic EE and finally holographic complexity would automatically acquire a dependence on the 't-Hooft coupling. To understand what is going on, we recall that the string background dual to the dipole-deformed CFT was obtained by performing a $TsT$ transformmation to the original (undeformed) AdS$_5\times$S$^5$ sugra background (refer to \cite{Bergman:2001rw} where this procedure is carried out in Sec. 4) with the factors of $R^2/\alpha' = s$ introduced in the resulting deformed NS-NS background. The deformed background can thus be thought of as being obtained after including (summing up) all order (classical) stringy/worldsheet corrections to the original background, just as in the boundary the deformation is represented by this irrelevant nonlocal dipole deformation term in the SYM Lagrangian (the nonlocal dipole term can be expanded to infinite number of local terms to the SYM lagrangian). The leading term in \eqref{DDSYMCA}, displays a volume law scaling and is independent of the dipole-deformation parameter $a$. This AdS result can be recovered by taking the limit $a \rightarrow 0$ of the CA expression \eqref{DDSYMCA}. However, notice that the full action complexity, also coincides with the pure AdS results when one sets $s=1$, even though the deformation $a=s \tilde{L}$ is not removed in this case! We believe this is an accidental agreement, and this agreement won't extend to other holographic observables such as correlators, Wilson loops and entanglement entropy. This is evident from the fact that setting $s=1$ in the gravity background Eq. (2.4)-(2.6) does not reduce to the undeformed $AdS_5\times S^5$ background. The universal (i.e. for all holographic observables) AdS result must be recovered only when the deformation parameter $a$ is set to zero. 

Asymptotically (small $\epsilon$),
\begin{align}
    C_{\mathcal{A}}\sim 4 \left(\frac{W}{\epsilon}\right)^3 \ln (3)-4 \left(\frac{W}{\epsilon}\right)^3 \left(s ^2-1\right)+\frac{12  W^3 }{a^2\epsilon}\left(s ^2-1\right)-\frac{6 \pi W^3}{a^3}\left(s ^2-1\right)
\end{align}
The behavior of action complexity in limiting regimes are as follows:
\begin{itemize}
\item In weak nonlocality regime,
\begin{equation}
    \mathcal{C}_{\mathcal{A}}=4 \left(\frac{W}{\epsilon}\right)^3 (\ln3-\ln\alpha+\ln l_{ct})+O\left(a^{2}\right)
\end{equation}
\item In strong nonlocality regime,
\begin{equation}
    \mathcal{C}_{\mathcal{A}}=  \frac{2}{3} \left(\frac{W}{\epsilon}\right)^3 \left(\alpha_{+}-6 \log (\alpha)-6 s ^2+6 \log (l_{ct})+5+6\log (3)\right)+O\left(\frac{1}{a}\right)\label{snl}
\end{equation}

From the limiting behavior of complexity in the weak nonlocality regime it can be seen that we get the AdS result either by fixing $\alpha=l_{ct}$ or $\alpha=1$ with $l_{ct}=R=1$. The strong nonlocality regime result \eqref{snl} is also worth reflecting on for a moment. For complexity to be positive definite, the t'Hooft coupling $s$ should satisfy the (upper) bound $\frac{l_{ct}+5}{6}+\ln(3)>s^{2}$, which depends expliciltly on $l_{ct}$ and reduce to $1+\ln(3)>s^{2}$ if we choose $l_{ct}=R=1$ . We speculate that this upper bound signals the breakdown of the UV-completeness of the DDSYM theory analogous to the LST case \cite{Katoch:2022hdf}.
\end{itemize}
\section{Noncommutative SYM theory: Anisotropic case} \label{NCSYM1}
In this section, we consider non-commutative super Yang-Mills (NCSYM) theory, which is a generalization of super Yang-Mills theory to a non-commutative space-time. We consider a maximally supersymmetric \(SU(N)\) non-commutative super Yang--Mills theory on \(\mathbb{R}^2_\theta \times \mathbb{R}^{1+1}\), where the non-commutative plane satisfies \([x,y] = i\theta\). We can use the following star product to define the non-commutative deformation of $\mathcal{N}=4$ SYM.
\begin{align}
    (f\star g)(x,y)=e^{\frac{i}{2}\theta\left(\frac{\partial}{\partial \xi_1}\frac{\partial}{\partial \zeta_2}-\frac{\partial}{\partial \xi_1}\frac{\partial}{\partial \zeta_2}\right)}f(x+\xi_1,y+\zeta_1)g(x+\xi_2,y+\zeta_2)|_{\xi_1=\zeta_1=\xi_2=\zeta_2=0}
\end{align}
Within the framework of the AdS/CFT correspondence, NCSYM emerges by considering the decoupling limit of D3-branes in the presence of NS--NS \(B\)-field. This setup was first explored in \cite{Hashimoto:1999ut, Maldacena:1999mh, Alishahiha:1999ci}, where it was shown that the dual gravitational background is no longer the standard AdS\(_5\times S^5\) but a deformed geometry. In particular, the metric takes the form
\[
ds^2 = \alpha'\left[U^2\Big(-dt^2+dx_1^2+h(U)(dx_2^2+dx_3^2)\Big)+\frac{1}{U^2}dU^2+d\Omega_5^2\right]\,,
\]
with
\[
h(U)=\frac{1}{1+a^4U^4}\,.
\]
Here, \(U\) is the radial holographic coordinate, and \(a\) is related to the noncommutative parameter \(\theta\) via the background \(B\)-field. This deformation encapsulates the anisotropic scaling induced by noncommutativity and directly influences the holographic dictionary.

The impact of noncommutativity on the dual field theory is profound. Two-point correlation functions exhibit a characteristic UV/IR mixing --- high energy (UV) behavior becomes entangled with long-distance (IR) physics, a signature of the underlying non-local interactions. Wilson loops, which in SYM serve as order parameters for confinement and phase transitions, acquire a dipole-like structure reflecting the noncommutative smearing of charge distributions. Furthermore, recent extensions of these analyses to quantum information theoretic measures, such as holographic entanglement entropy and computational complexity \cite{Karczmarek:2013xxa}, reveal modified scaling laws and phase structures. Also see \cite{Fischler:2013gsa, Couch:2017yil, Fischler:2018kwt, Eccles:2021zum} for related work on aspects of entanglement entropy, complexity, information and entanglement spreading, and chaos and operator growth in the anisotropic NCSYM in the holographic context. 
\subsection{Volume complexity for the anisotropic case}
The Holographic dual background \cite{Karczmarek:2013xxa} dual to non commutative gauge theory on  $\mathbb{R}_{\theta}^2\times \mathbb{R}^{1+1}$ is
\begin{align}\label{NCSYM}
    ds^2&= u^2(-dt^2+h(u)(dx^2+dy^2)+dz^2)+ \frac{du^2}{u^2}+ d\Omega_5^2\nonumber\\
    e^{2\phi}&=g_s^2h(u)\nonumber\\
    B_{xy}&=-\frac{1-h(u)}{\theta}=-\frac{1}{\alpha'}a^2u^4h(u)\nonumber\\
    h(u)&=\frac{1}{1+(a u)^4}  
\end{align}
Where, length scale of noncommutativity is $a =s^{1/4}\sqrt{\theta}$,  $s $ is (dimensionless) t'Hooft coupling, $\sqrt{\theta}$ is the noncommutativity parameter and $B_{xy}$ is the only non-zero component of the NS-NS form background. The volume complexity can be found by taking a space like hypersurface specified by the embedding $t=t(u)$ in the string frame as,
\begin{align}
     dh^2&=\left(\frac{1 }{u^2}- u^2t'(u)^2\right)du^2+  u^2h(u)(dx^2+dy^2)+ u^2dz^2+ d\Omega_5^2
\end{align}
The volume of the hypersurface with the correct factors accounted for being in the string frame is 
\begin{align}
    V(t_*)&=\int d^{9}\sigma  ~e^{-2(\phi-\phi_0)}\sqrt{G}=R^9\,W^3\,\omega_5\,\int_{u_*}^{u_b} du\, u^2\sqrt{1-u^4 t'(u)^2},\label{V}
\end{align}
Here $W$ is an IR cutoff in the boundary $-W/2\leq x,y,z \leq W/2$. By inspection of (\ref{V}), it is evident that the global maxima occurs when $t'(u)=0$.  implying that the volume complexity is 
\begin{align}
    \mathcal{C}_V(t_*)&=\frac{\omega_5 }{3 G_N^{(10)}}\, W^3 u_b^3
\end{align}
Here again we re-express the final result by replacing the bulk UV cutoff in terms of the boundary lattice spacing $\epsilon$ via $u_b = 1/\epsilon$) and central charge $c$.
\begin{align}
    \mathcal{C}_V(t_*)&=\frac{1 }{3 G_N^{(5)}}\, \left(\frac{W}{\epsilon}\right)^3=\frac{8c}{3\pi}\left(\frac{W}{\epsilon}\right)^3
\end{align}
Again, similar to the dipole case, the volume complexity is independent of noncommutative parameter $a$. This could have been predicted from the volume functional \eqref{V}, where the noncommutative deformation factors from the metric coefficients of the two noncommutative directions, and the deformation factors arising from the dilaton profile cancels each other, even before maximization, a purely dimensional accident or conspiracy. In terms of the field theory picture, one can interpret this in the following way: the loss of degrees of freedom due to noncommutativty in the UV is exactly compensated by the usual addition of degrees of freedom as one RG-flows from IR to the UV. As we will shortly see this dimensional conspiracy is evaded when we consider the fully isotropic NCSYM theory.\\

\subsection{Subregion Volume complexity (CV) for the anisotropic case}
To compute subregion complexity we
focus our attention on the portion of the maximal volume slice which is contained within the Ryu-Takayanagi (RT) surface, and which is also homologous to the boundary subregion of interest. We take the boundary subregion to be the infinitely long strip, $-\frac{l}{2}<x<\frac{l}{2},\  -\frac{W}{2}<y,z<\frac{W}{2}$ in the limit $W\to \infty$ with $l$ fixed. In this case the pullback metric on the bounding codimension 2 (RT) surface is 
\begin{align}\label{NCSYM}
    d\gamma^2&= u(x)^2h(u)dy^2+u(x)^2dz^2+u(x)^2h(u)\left(\frac{u'(x)^2}{h(u)u(x)^4}+ 1\right)dx^2+ d\Omega_5^2
    \end{align}
     By specifying the bulk minimal area surface as $u= u(x)$ the area is given by
    \begin{align}
    \mathcal{A}&=\omega_5 W^2 \int_{-l/2}^{l/2}dx\, u(x)^3 \sqrt{1+\frac{u'(x)^2}{h(u)u(x)^4}}\label{subNCYM}
    \end{align}
    This is the typical minimax variational problem that we encounter in the problems in dynamics with $x$ playing the role of time with the lagrangian given by
    \begin{align}
        \mathcal{L}(u(x),u'(x))&=u(x)^3 \sqrt{1+\frac{u'(x)^2}{h(u)u(x)^4}}
    \end{align}
    Since the lagrangian doesn't depend explicitly on time, we can construct a Hamiltonian and 
    evaluate at the boundary using the boundary conditions at the turning point $u_*$ occurring at $x=0$ where $u'(0)=0$ and with this input, the constant value of the Hamiltonian is $\mathcal{H}(u_*)=-u_*^3$. To obtain 
    \begin{align}
        u'(x)=\frac{u^2 \sqrt{h(u)} \sqrt{u^6-u_*^6}}{u_*^3}
    \end{align}
The length of the subregion is given by,
    \begin{align}
        \int_{0}^{l/2}dx&=\int_{u_*}^{u_b}\frac{du}{u'}\label{subsize}\nonumber\\
     \Rightarrow l   &=2 u_*^3\int_{u_*}^{u_b}\frac{ du}{u^2 \sqrt{h(u)} \sqrt{u^6-u_*^6}}
    \end{align}
Since the above integration is not analytically tractable in general, and to gain traction we look at two different regimes, namely near AdS (large subregions) and far from AdS (small subregions) to obtain $l$ in terms of $u_*$, which will be useful to obtain complexity in terms of subregion length $l$ in later subsections.\\

In the near AdS regime ($au\ll 1$), we find $l$  in terms of $u_*,a$ by redefining new variables as $\overline{a}=a/l$, $x_b=l u_b$ and $x_*=l u_*$,
\begin{align}
    x_*=&2x_*^4\int_{x_*}^{x_b}\frac{dx\sqrt{1+\bar{a}^4 x^4}}{x^2\sqrt{x^6-x_*^6}}=2x_*^4\int_{x_*}^{x_b}\frac{dx}{x^2\sqrt{x^6-x_*^6}}\left(1+\frac{\bar{a}^4x^4}{2}+O(\bar{a}^{8})\right)\nonumber\\
    x_*\sim & \frac{2 \sqrt{\pi } \Gamma \left(\frac{2}{3}\right)}{\Gamma \left(\frac{1}{6}\right)}+\frac{1}{3} \overline{a}^4 x_*^4 \ln \left(\frac{2x_b^3}{x_*^3}\right)+O(\bar{a}^{8})
\end{align}
We can invert to write $l$ in terms of $u_*$ by solving perturbatively in $\overline{a}$ to get,
\begin{align}
     u_*&=\frac{1}{l}\frac{2 \sqrt{\pi } \Gamma \left(\frac{2}{3}\right)}{ \Gamma \left(\frac{1}{6}\right)}+\frac{a^8}{l^9}\frac{128 \pi ^{7/2}  \Gamma \left(\frac{2}{3}\right)^7}{ \Gamma \left(\frac{1}{6}\right)^7} \ln ^2\left(\frac{2^{2/3} \sqrt{\pi } \Gamma \left(\frac{2}{3}\right)}{l u_b \Gamma \left(\frac{1}{6}\right)}\right)+O\left(a^{16}\right)
\end{align}\label{ustarnearads}
Similar to near AdS case, we can also work out $l$ in terms of $u_*$ in the far from AdS regime as,
\begin{align}
   x_*&=2x_*^4\int_{x_*}^{x_b}\frac{dx\  \overline{a}^2x^2}{x^2\sqrt{x^6-x_*^6}}\left(1+\frac{1}{2\overline{a}^4 x^4}+\dots\right)\nonumber\\
   x_*&\sim -\frac{\Gamma(-\frac{1}{6})}{2\sqrt{\pi}\overline{a}^2 \Gamma(\frac{1}{3})}\label{uStarinlfarads}\\
    u_*&\sim \frac{l}{a^2}\frac{\Gamma(\frac{5}{6})}{2\sqrt{\pi} \Gamma(\frac{4}{3})}
\end{align}
Now we can find entanglement entropy by plugging (\ref{subsize}) back in (\ref{subNCYM}) to obtain, 
\begin{align}
    \mathcal{A}&= 2  \omega_5 W^2 \, \int_{u_*}^{u_b}du\,\frac{ u^4}{ \sqrt{h(u)} \sqrt{u^6-u_*^6}}
\end{align}
    The result was previously obtained in \cite{Barbon:2008ut, Karczmarek:2013xxa}. We however, want to go ahead and be able to compute the subregion complexity given by the codimension-1 surface in the bulk bounded by the RT curve. The pullback of the ambient metric on the $t'(u)=0$ foliation of the embedding surface is 
    \begin{align}
        dh^2&= u^2(h(u)(dx^2+dy^2)+dz^2)+ \frac{du^2}{u^2}+ d\Omega_5^2
    \end{align}
    We use (\ref{V}) to find the volume of the hypersurface and the volume complexity as,
    \begin{align}
        \mathcal{V}&=2\omega_5\int_{-W/2}^{W/2}dy\int_{-W/2}^{W/2}dz\,\int_{u_*}^{u_b}du\,\int_{0}^{x(U)}dx\, h(u)\, u^2\,e^{-2(\phi-\phi_0)}\\
        &\mathcal{C}_V=\frac{2W^2u_*^3\omega_5}{G_N^{(10)}}\,\int_{u_*}^{u_b}du\,u^2\,\int_{u_*}^{u}\frac{ d\Tilde{u}}{\Tilde{u}^2 \sqrt{h(\Tilde{u})} \sqrt{\Tilde{u}^6-u_*^6}} \label{sub aniso}
    \end{align}
The above integral is not analytically tractable, so we {plot the numerical integration results as shown in figure {\ref{fig:LogCVvslogl}} for small values of the noncommutative deformation parameter (for large values of the deformation the numerics are not quite reliable or stable). The value of the critical subregion size (length) for three different values of $a$ extracted from the plots (in units of AdS radius) are as displayed in the table below. We also provide analytical estimates of the critical subregion size for comparison.
\begin{table}[H]
    \centering
    \begin{tabular}{|c|c|c|}
    \hline
    $a$ & Analytical value & Value from Plot\\
    \hline
    $10^{-9}$&  $1.74\times 10^{-9}$ & $1.64\times 10^{-9}$\\
    \hline
    $10^{-7}$&  $1.74\times 10^{-7}$ & $1.64\times 10^{-7}$\\
    \hline
    $10^{-5}$& $1.7\times 10^{-5}$   & $1.6\times 10^{-5}$\\
    \hline
    $10^{-3}$& 0.0017  &0.0016\\ 
    \hline
    \end{tabular}
\end{table}
From the log-log plot one can see for large subregion size, the subregion complexity scales linearly with subregion size (length) while for small subregion size the subregion complexity scales cubically with the subregion length.
\begin{figure}
    \centering
    \includegraphics[width=0.4\linewidth] {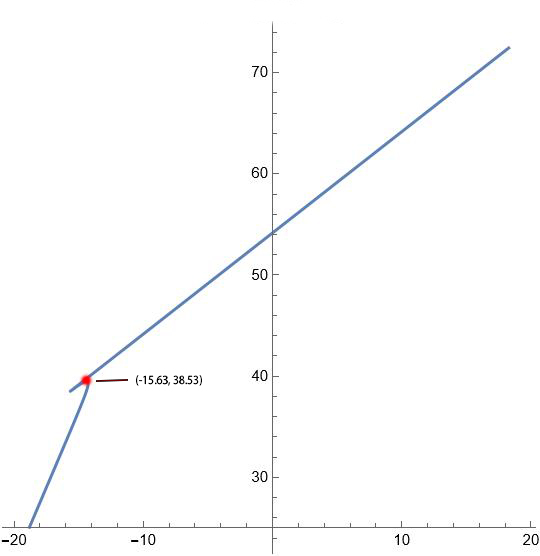} \hspace{0.1cm} 
    \includegraphics[width=0.4\linewidth]{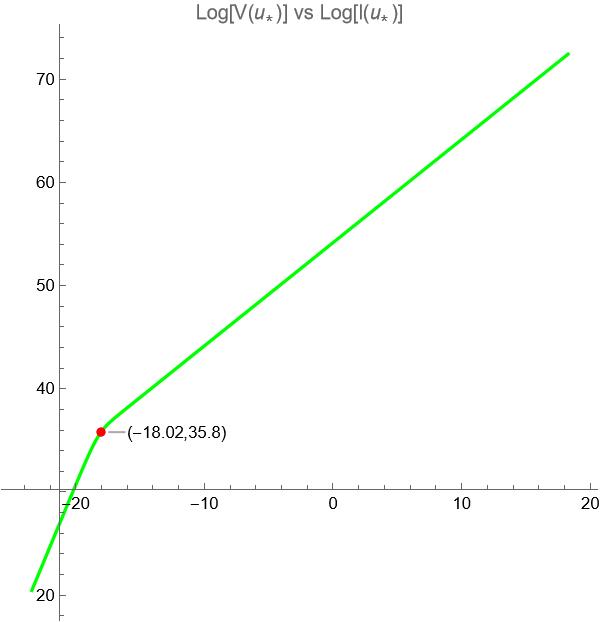}    \includegraphics[width=0.4\linewidth]{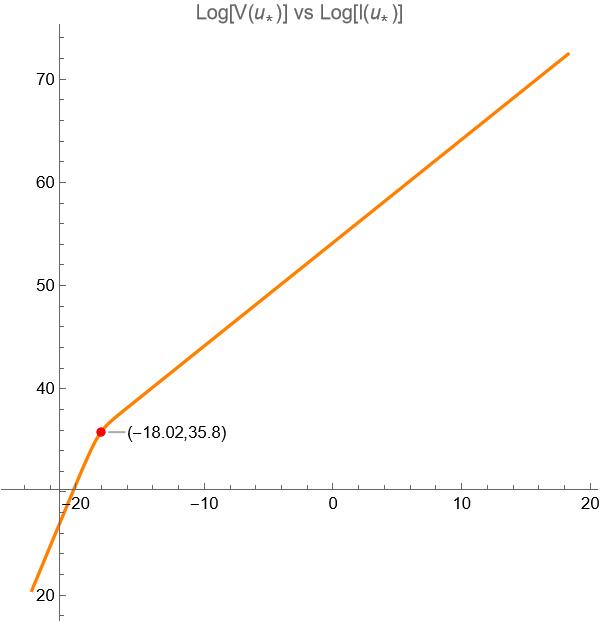}   \hspace{0.1cm}  \includegraphics[width=0.4\linewidth]{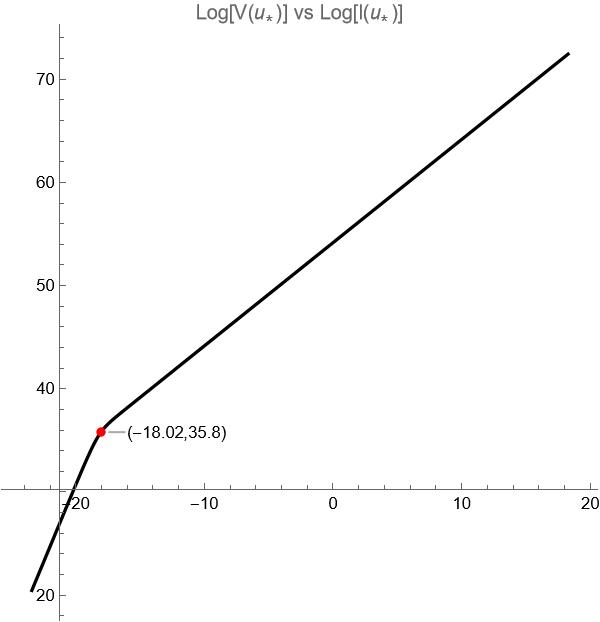} 
    \includegraphics[width=0.4\linewidth]  {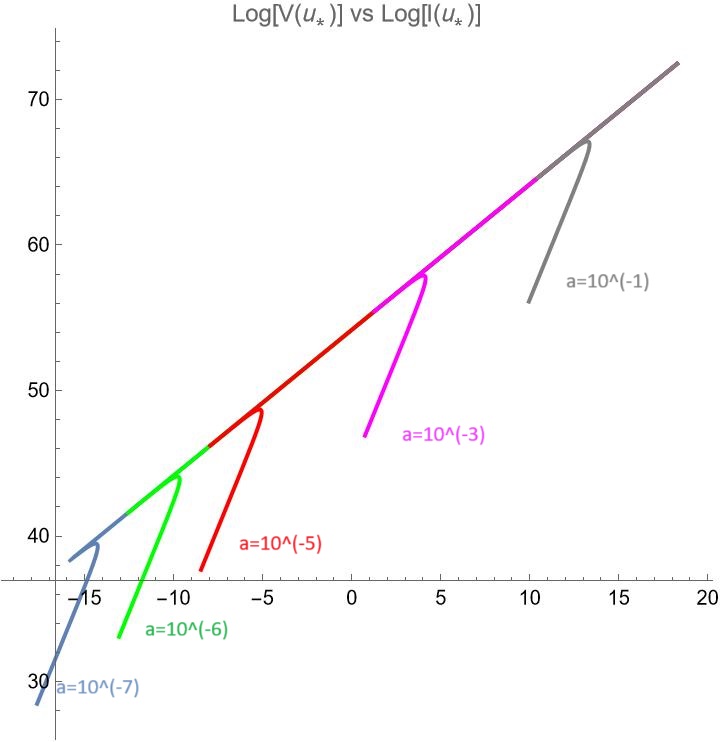} 
    \caption{Numeric plots of $\log(\mathcal{C}_V)$ vs $\log(l)$ for different values of $a=10^{-7}, 10^{-9}, 10^{-11}, 10^{-13}$ respectively. The transition points are indicated in respective plots.} \label{fig:LogCVvslogl}
\end{figure}

Next, we analytically work out the complexity functional \eqref{sub aniso} in two extreme regimes, specifically near AdS or weakly noncommutative case and the extremely noncommutative or far from AdS case.
\subsubsection{Near AdS/Weakly noncommutative regime}
In the near AdS regime ($\bar{a}\ll 1$), the integrand simplifies and hence we can find the Volume complexity as,
\begin{align}
    V&=W^2  \frac{2x_*^3}{l^2}\int_{x_*}^{x_b} x^2 dx\int_{x_*}^{x}\frac{1}{\tilde{x}^2\sqrt{\tilde{x}^6-x_*^6}}\left(1+\frac{\overline{a}^4 x^4}{2}+O(\bar{a}^{8})\right)\nonumber\\
    &=W^2\left(\frac{\sqrt{\pi }}{81 u_*} \left(\frac{2 \sqrt{3} \pi  u_b^3}{\Gamma \left(\frac{7}{6}\right) \Gamma \left(\frac{4}{3}\right)}-\frac{9 u_*^3 \Gamma \left(\frac{1}{6}\right)}{\Gamma \left(\frac{2}{3}\right)}\right)+\frac{1}{9} a^4 u_b^3 u_*^3 \left(-3 \ln \left(\frac{u_*}{\sqrt[3]{2} u_b}\right)-1\right)+O(a^8)\right)\label{VinuStarnearads}
\end{align}
We can find the Volume complexity in terms of subregion length by substituting (\ref{ustarnearads}) in (\ref{VinuStarnearads}) to get,
\begin{align}   \mathcal{C}_V&=\frac{ W^2}{G_N^{(5)}}\left(\frac{l }{3 \epsilon^3}-\frac{4 \pi ^{3/2} \Gamma \left(\frac{2}{3}\right)}{9 l^2 \Gamma \left(\frac{1}{6}\right)}+a^4 \left(\frac{128 \pi ^{9/2} \epsilon^3 \Gamma \left(\frac{2}{3}\right)^9}{9 l^9  \Gamma \left(\frac{1}{6}\right)^9}-\frac{8 \pi ^{3/2}  \Gamma \left(\frac{2}{3}\right)^3}{9 l^3 \epsilon^3\Gamma \left(\frac{1}{6}\right)^3}\right)+O(a^8)\right)
\end{align}
As a consistency check, upon substituting $a=0$ we recover the commutative SYM (pure AdS) result \cite{Ben-Ami:2016qex}. 
\subsubsection{Far AdS/extreme noncommutative regime}
Similar to near AdS regime, if we take the far AdS regime($\bar{a}\gg 1$), the volume complexity can be found as,
\begin{align}
    V&=W^2  \frac{2x_*^3}{l^2}\int_{x_*}^{x_b} x^2 dx\int_{x_*}^{x}d\tilde{x}\frac{\overline{a}^2}{\sqrt{\tilde{x}^6-x_*^6}}\left(1+\frac{1}{2\overline{a}^4x^4}-\frac{1}{8 \overline{a}^8 u^8}+O\left(\frac{1}{\overline{a}^{12}}\right)\right)
\end{align}
We can do the above integration and substitute (\ref{uStarinlfarads}) to write the Volume Complexity in terms of $l$ as follows,
\begin{align}
    \mathcal{C}_V&=\frac{1}{G_N^{(5)}}\left(\frac{W^2l }{3\epsilon^3}+\frac{W^2l^4}{a^6}\frac{\sqrt{3} \Gamma \left(\frac{5}{6}\right)^6}{\sqrt[3]{2} \pi ^3\Gamma \left(\frac{4}{3}\right)^3}-\frac{W^2l^3 }{a^4\epsilon}\frac{ \Gamma \left(-\frac{1}{6}\right)^4}{32 \sqrt[3]{2} \sqrt{3} \pi ^3 \Gamma \left(\frac{1}{3}\right)}\right)
\end{align}
Surprisingly, the leading divergence piece is independent of the noncommutativity parameter $a$, and identical to the leading piece of the commutative. So the volume law scaling (or equivalently linear in extension $l$ along the noncommutative deformation)  is retained even in the nonlocal (noncommutative) domain. We speculate about the potential causes for this result in Sec. \ref{DO}. The subleading divergence and the finite piece evidently defies volume law scaling (nonlinear in $l$), the linear divergence displays a cubic scaling law and the finite piece scales quartically with $l$. \\
\subsection{Action complexity (CA) for the anisotropic case (10 D) }
One attractive feature of noncommutativity  is that it encodes nonlocality into the theory that we wish to probe using holographic complexity.  With this in sight, we examine geometries derived from string theory by turning on the NS-NS $B$-field on Dp-branes. The presence of a nonzero $B$-field imposes Dirichlet boundary conditions on open strings, leading to a nonvanishing commutator between the coordinates at their endpoints. Upon decoupling closed strings, the Dp-brane's worldvolume effectively becomes a noncommutative space.\\

 The low energy limit of Type IIB string theory reduces to  the following SUGRA bulk action \cite{Couch:2017yil} with Neveu-Schwartz, Ramond-Ramond and Chern-Simmons sector outlined below
   \begin{align}
    S_{\mathcal{V}}&=S_{NS}+S_{R}+S_{CS}\nonumber,\\
    \text{where,}\hspace{1cm }S_{NS}&=\frac{1}{2\kappa_{10}^2}\int d^{10}X \sqrt{-g}e^{-2\Phi}\left(\mathcal{R}+4\partial_{\mu}\Phi\partial^{\mu}\Phi-\frac{1}{2}|H_3|^2\right)\nonumber,\\
    S_{RR}&=-\frac{1}{4\kappa_{10}^2}\int d^{10}X\left(|F_1|^2+|\Tilde{F}_3|^2+\frac{1}{2}|\Tilde{F}_5|^2\right)\nonumber,\\
    S_{CS}&=-\frac{1}{4\kappa_{10}^2}\int C_4\wedge H_3 \wedge F_3\nonumber.
\end{align}
Where, $\Phi$ is dilaton, $H_3$ is the field strength of NS 2-form $B_2$, $F_p$ is the field strength of RR (p-1) forms $C_{p-1}$ and fluxes
\begin{align}
    H_3&=dB_2, & \Tilde{F}_3&=F_3-C_0H_3,\nonumber\\
    F_p&=dC_{p-1}, & \Tilde{F}_5&=F_5-\frac{1}{2}C_2\wedge H_3+\frac{1}{2}B_2\wedge F_3,\nonumber.
\end{align}
 The background field content spans the following form fields 
\begin{align}
    e^{-2\Phi}&=\frac{1}{g_s^2h(r)}, & B_{(2)}\equiv B_{12}&=- a^2 u^4h(u),\nonumber \\
   C_{(2)}\equiv C_{01}&=\frac{a^2}{g_s}u^4, & F_{(5)}\equiv F_{0123u}&=\frac{4}{g_s}u^3h(u),\nonumber\\
 C_{(0)}&=0.\nonumber
\end{align}

Furthermore,  the Type-IIB action is supplemented with the self-duality condition $\Tilde{F}_5=\star \Tilde{F}_5$. \\
Holographic dual background \cite{Karczmarek:2013xxa} dual to non commutative gauge theory on  $\mathbb{R}_{\theta}^2\times \mathbb{R}^{1+1}$ is
\begin{align}\label{NC-SYM}
    ds^2&=u^2(-dt^2+h(u)(dx^2+dy^2)+dz^2)+R^2 \frac{du^2}{u^2}+ d\Omega_5^2\nonumber,\\
      h(u)&=\frac{1}{1+(a u)^4} \nonumber, \\
    e^{2\phi}&=g_s^2h(u),\nonumber\\
    \quad B_{xy}&=-\frac{1-h(u)}{\theta}=-a^2u^4h(u).
   \end{align}
Where, we have normalized the AdS radius to unity. 
The expression for the Ricci scalar is given by
\begin{align}
      \mathcal{R}&=\frac{8 a^4 u^4 \left(2 a^4 u^4+9\right)}{ \left(a^4 u^4+1\right)^2}\nonumber.
\end{align}
${t,x_i}$ are the D3 brane coordinates, while ${x,y}$ are non-commuting coordinates spanning the non-commutative plane $\mathbb{R}_{\theta}^2$ with Moyal algebra $[x,y]=i\theta$.

As previously discussed, the Wheeler-DeWitt (WdW) patch defines the domain of integration for the bulk action and is enclosed by null hypersurface
\begin{align}
     t_+(u)-t_-(u) &=2\int^{u_b}_{u}\frac{du'}{u'^2}\nonumber.
\end{align}

The gravitational action receives contributions from several boundary terms of varying codimensions. We initiate our analysis by computing the codimension-0 contribution, which corresponds to the bulk term.
\begin{itemize}
       \item \textbf{Evaluation of the bulk action:} 
       The theory has been obtained by the nonlocal irrelevant deformation of $\mathcal{N}=4$ SYM which distorts the theory in the UV to non-commutative super Yang-Mills, leaving the gravitational dual AdS intact in the IR. This is also visible from the dimensionality of the coupling constant $a$ which has inverse energy or length dimensions and therefore, we expect the subleading terms to be more and more divergent in the UV. Hence, we consider two perturbative regimes in the following. Firstly, where the non-commutativity is weak, and we have $a\to 0$ and $  u_b\to \infty $ in such a manner that the combination $au_b\sim 0.$ Secondly, we work in the strongly non-commutative regime where non-commutative scale $a\simeq 1$ while $u_b\to \infty. $\\
       
    The bulk term is computed over the WdW patch, where the integration measure is given by $\sqrt{-g} = u^3 h(u)$. We list all the contributions coming from the different sectors of SUGRA action below.
       \item  \textbf{Einstein-Hilbert term\footnote{Here we use $$\frac{\omega_5}{2\kappa^2_{10}}=\frac{\omega_5}{16\pi G_N^{(10)}}=\frac{1}{16\pi G_N^{(5)}}$$ }} 
        \begin{align}
           S_{EH}&=\frac{1}{2\kappa_{10}^2}\int d^{10}X\,\sqrt{-g}e^{-2(\phi-\phi_0)}\mathcal{R},\nonumber\\
           &=\frac{1}{16\pi G_N}\int^{u_b}_0 du\, 2 L^3 R^5 u^3 \left(\frac{1}{u}-\frac{1}{u_b}\right) \left(\frac{8 a^4 u^4 \left(2 a^4 u^4+9\right)}{R^2\left(a^4  u^4+1\right)^2}\right) .\, \nonumber
           \end{align}
        In the regime when $a\ll \frac{1}{u_b}$, which we will henceforth refer to as  near AdS regime,
        \begin{align}
            S_{EH}&=\frac{L^3u_b^3}{16\pi G_N}\left(\frac{18}{7} a^4u_b^4\right)+O(a^{8}).\nonumber
        \end{align}
        In this regime, Einstein-Hilbert part of the action does not receive any leading AdS contribution.
  In the regime when $a>>\frac{1}{u_b}$, that we will refer to as the  strong non-commutative regime, 
        \begin{align}
            S_{EH}&=\frac{L^3  u_b^3}{16\pi G_N}\frac{8}{3} +O\left({a^{-3}}\right).\label{aiso-EH}
        \end{align}
      \item  \textbf{Dilaton Kinetic term}   
      \begin{align}
  S_{\phi}&=\frac{8\omega_5L^3}{2\kappa^2_{10}}     \int_{0}^{u_b}  du\,g^{uu}(\partial_{u}\phi)^2 u^3\int^{u_b}_{u}\frac{du'}{u'^2},\nonumber\\
    &=\frac{32 a^8 L^3}{16\pi G_N}\int_0^{u_b} du\,\frac{u^{11}}{\left(a^4 u^4+1\right)^2} \left(\frac{1}{u}-\frac{1}{u_b}\right) .\nonumber
\end{align}
In the near AdS regime,
\begin{align}
    S_{\phi}=\frac{L^3u_b^3}{16\pi G_N}\left(\frac{8}{33} a^8u_b^{8}\right)+O(a^{12}).\nonumber
\end{align}
The dilaton piece, contributes only at the sub-sub leading order. 
Whereas, in the strong non-commutative regime,
\begin{align}
  S_{\phi}&=\frac{L^3u_b^3}{16\pi G_N}   \frac{8}{3} +O\left({a^{-3}}\right).\label{aiso-dilaton}
\end{align}
\item \textbf{Neveu-Schwarz B field term}
\begin{align}
    S_{B}&=\frac{-1}{2\kappa_{10}^2}\int d^{10}X\,\sqrt{-g}e^{-2(\phi-\phi_0)}\left(\frac{H_{uxy}H^{uxy}}{2}\right),\nonumber\\
&=\frac{32 L^5a^4 L^3 }{16\pi G_N}\int_0^{u_b}\frac{u^7}{\left(a^4 u^4+1\right)^2}\left(\frac{1}{u}-\frac{1}{u_b}\right).\nonumber
\end{align}
In the near AdS regime, 
\begin{align}
    S_{B}=\frac{L^3u_b^3}{16\pi G_N}\left(\frac{4 a^4 u_b^4 }{7 }\right)+O(a^{8}).\nonumber
\end{align}
In the strong noncommutative regime,
\begin{align}
    S_{B}&=O\left(a^{-3}\right).\label{aiso-B}
\end{align}
 \item \textbf{RR sector: Ramond-Ramond gauge field }
\begin{align}
     S_{RR}&=-\frac{1}{4\kappa_{10}^2}\int d^{10}X|\Tilde{F}_3|^2,\nonumber\\
     &=\frac{16\omega_5 a^4 L^3 }{2
    \kappa_{10}^2}\int _0^{u_b}du \frac{u^7 }{(a^4 u^4+1)}\left(\frac{1}{u}-\frac{1}{u_b}\right).\nonumber
\end{align}
In the near AdS regime, 
\begin{align}
    S_{RR}=\frac{L^3u_b^3}{16\pi G_N
    }\left(\frac{2}{7} a^4 u_b^4\right)+O(a^{8}).\nonumber
\end{align}
In the strong noncommutative regime 
\begin{align}
    S_{RR}&=\frac{1}{16\pi G_N}\left(\frac{4}{3}  L^3 u_b^3\right)+O\left(a^{-3   }\right).\label{aiso-RR}
\end{align}

\item \textbf{Topological boundary counterterm}\\
Like we mentioned in the previous section, it  turns out that we need to incorporate the appropriate topological boundary  term \cite{Kurlyand:2022vzv,Apolo:2025wcl} to get correct action leading to the gravitational background with a negative cosmological constant. 
 \begin{align}
    F_5&=F_5^{(el)}+F_5^{(mag)} ,\nonumber\\
     F_5^{(mag)}&=\star F_5^{(el)},\nonumber\\
       \text{Where, } \,\,\,F_{(5)}^{(el)}&\equiv F_{0123u}=4u^3h(u) \,dt\wedge dx\wedge dy\wedge dz\wedge du ,\nonumber\\
       \text{and, }\,\,\,\, F_5^{(mag)}&=4d\theta_1 \wedge d\theta_2\wedge....\wedge d\theta_5.\nonumber
    \end{align}
 The individual contribution of each of these terms can be seen to be factored into the flux over the AdS part and the sphere $S^5$ part in the following manner 
  \begin{align}
        S_{top}&=\gamma \int F_{5AdS}\wedge F_{5S^5}\nonumber,\\
        &=-\frac{1}{4(5!)^2\kappa_{10}^2}\int F_5^{(el)}\wedge F_5^{(mag)}\nonumber.
    \end{align}
    Following this prescription, the corresponding action can be seen to be of the following form
  \begin{align}
        S_{top}
        &=-\frac{1}{4\kappa_{10}^2}\int_0^{u_b} du \,16u^3h(u)\int dtdxdydz\int d\omega_5\nonumber,\\
        &=-\frac{8}{16\pi G_N}\int dxdydz\int_0^{u_b}du \frac{u^3 }{1+a^4u^4}\left(
        \frac{1}{u}-\frac{1}{u_b}\right)\nonumber.
    \end{align}
    Upon tuning the couplings accordingly, we recover in the near AdS regime
    \begin{align}
          S_{top}  &=\frac{L^3u_b^3}{16\pi G_N}\left( -\frac{4}{3}\right)+O(a^{4})\nonumber.
    \end{align}
    And the strongly non-commutative regime
    \begin{align}
         S_{top}&=O\left(a^{-3}\right).\nonumber
    \end{align}
Combining all the contributions, corresponding to the near AdS regime, reduces to the following perturbative expansion in terms of non-commutative coupling 
\begin{align}
  S_{\mathcal{V}}   =\frac{L^3u_b^3}{16\pi G_N}\left(-\frac{4}{3}\right)+O(a^{-4}). \label{aniso_AdSbulk}
\end{align}
 Before incorporating non-commutative effects, we first validate the expected AdS result. Specifically, evaluating the bulk action in the perturbative regime where the UV cutoff scale is much larger than the non-commutative coupling, i.e., $a \ll 1/u_b$, correctly reproduces the standard AdS behaviour \cite{Reynolds:2016rvl}.
On the other hand, in the strongly non-commutative regime when $au_b\gg 1$, the bulk contribution is given by summing \eqref{aiso-EH}, \eqref{aiso-dilaton}, \eqref{aiso-B} and  \eqref{aiso-RR}  to be 
\begin{align}
     S_{\mathcal{V}}&=\frac{L^3u_b^3}{16\pi G_N}\left(\frac{20}{3}  \right)+O\left(a^{-3}\right). \label{aiso-bulk} 
\end{align}
   A detailed discussion on the scaling behaviour of different terms will follow once all contributions are accounted for. We now proceed to evaluate the boundary term which originates from the null boundaries of the WdW patch. 
\end{itemize}

\begin{itemize}
    \item \textbf{Contributions from null boundaries:} The gravitational action receives contributions from summing over all null boundary components, given by
      \begin{align}
            S_{\partial\mathcal{V} }&=\sum_{j=\pm}\frac{2}{16\pi G_N}\int d^3x\int ds_{j}\,\sqrt{\gamma_{j}}\,\kappa_{j}.\label{aniso-null}
        \end{align}
    Here, the affine parameter $s$ along the null generators increases in the future direction, and $\gamma$ represents the induced metric on constant-$s$ slices. On the section of constant $s_{\pm}=\mp u$,
\begin{align}
d\gamma^2= u^2(h(u)(dx^2+dy^2)+dz^2)+d\Omega_5^2\nonumber.
    \end{align}
    Additionally, the null normal vector satisfies $n^{\mu}\nabla_{\mu} n^{\nu} = \kappa n^{\nu}$.
    The boundaries of the WdW patch consists of null surfaces $\mathcal{N}_{\pm}$ 
    \begin{align}
      (t-T)^2 &=\left(\frac{1}{u}-\frac{1}{u_b}\right)^2.\nonumber
  \end{align}
    These null hypersurfaces, $\mathcal{N}_{\pm}$ are given by
    \begin{align}
      t &=T\mp\left(\frac{1}{u}-\frac{1}{u_b}\right),\nonumber
  \end{align}
    where the $+$ and $-$ indices denote the upper and lower null boundaries, respectively. The affine parameters are chosen such that $s_{\pm} = \mp u$, leading to the following null normal vectors
    \begin{align}
    n_{\pm}&=\left(\frac{\alpha}{u^2}, \mp\alpha 
    , \vec{0}\right), \label{anisonull}
\end{align}
    where, $\alpha$ is the normalization constant which cannot yet be uniquely fixed due to the normals being null. 
    This choice determines the parameter $\tilde\kappa$ from the relation
    \begin{align}
        n\cdot \nabla n^{\mu}_{\pm}&={\kappa}_{\pm}n^{\mu}_{\pm}.\nonumber
    \end{align}
     But one must keep in mind that owing to the fact that we are working with the string frame metric, the parameter $\tilde\kappa$ also scales differently under conformal transformation, which now reads 
  \begin{align}
      {\kappa}&=\tilde\kappa-\frac{1}{2} n^{\mu}\nabla_{\mu}\ln \phi. \nonumber
  \end{align}
Where, $\tilde\kappa$ above refers simply to its expression in Einstein frame metric and the conformal factor relates Einstein frame metric to string frame metric via $\tilde{g}_{\mu\nu}=e^{-\frac{\phi}{2}}g_{\mu\nu}$. It turns out that the above choice of the null parametrization is affine in Einstein frame, resulting in $\kappa=0$ leaving us with 
\begin{align}
    \kappa_{\pm}&=\pm\frac{a^4\alpha    u^3}{a^4 u^4+1},\nonumber
\end{align}
    Summing over both null boundary contributions and substituting into Eq.~(\ref{aniso-null}), we obtain:\\
   \begin{align}
       S_{\partial\mathcal{V}}&=-\frac{4a^4\alpha  L^3}{16 \pi G_N}\int_0^{u_b} du \left(\frac{ u^6}{a^4 u^4+1}\right).\nonumber
   \end{align}
    In the perturbative limit where $a \ll 1/u_b$, this expression receives vanishing contribution up to  $\mathcal{O}(a^4)$ in non-commutative coupling. This is  consistent with the the expected AdS result. \\

    In contrast, the strongly non-commutative regime (where $a \gg 1/u_b$) contribution is given by
    \begin{align}
     S_{\partial \mathcal{V}}&=\frac{L^3u_b^3}{16\pi G_N}\left(-\frac{4}{3}\alpha  \right)+O\left(\frac{1}{a^3}\right).\label{aiso-bound}
\end{align}
\end{itemize}

\begin{itemize}
    \item \textbf{LMPS counterterm prescription:} 
    As noted in \cite{Lehner:2016vdi}, the gravitational action depends on the choice of the affine parameter $s$ on null boundaries. To preserve reparametrization invariance, the following counterterm must be included for each null boundary\begin{align}
   \sum_{\pm}\Delta S_i&= -2 \int d{A_i} \,ds_i \,{\Theta_i} \ln| l_{ct}{\Theta_i}|, \hspace{3cm} \left({\Theta_i}\equiv \frac{1}{\sqrt{{\gamma}}}\frac{\partial \sqrt{{\gamma}}}{\partial s_i}\right).\nonumber
\end{align}
    
    Here, ${\gamma}$ denotes the induced metric on constant-$s$ cross-sections, which form codimension-2 surfaces. On slices with constant $s_{\pm} = \mp r$, the integration measure is as derived earlier ($ \sqrt{\gamma}=u^3h(u)$).
    After accounting for sign conventions ($\Theta_+ = -\Theta_-$), and to the causal structure, the total contribution from both surfaces turns out to be 
    \begin{align}
        \Delta S&=\frac{4}{16\pi G_N}\int d{A}\,\int_{0}^{\infty} ds_- \,{\Theta}_- \ln|{l_{ct}\Theta}_-|,\\\nonumber\  &=\frac{12L^3}{16\pi G_N} \int^{u_b}_{0} d  u \,u^ 2\ln\left|\frac{3l_{ct}}{u}\right|.\nonumber
    \end{align}
   The dimensionful parameter $l_{ct}$ is required in the  argument of logarithm to make it appear dimensionless.  Counterterm action can be evaluated to give 
    \begin{align}
         \Delta S  &=\frac{L^3u_b^3 }{16\pi G_N}\left(\frac{4}{3}+ 4 \ln \left(\frac{3l_{ct}}{ u_b}\right)\right).\label{aniso-ct}
    \end{align}
   Due to the absence of the noncommutative coupling parameter $a$, this the result for the counterterm action in the near AdS regime.  In the end, the consistency with the known AdS result  will enable us to  fix  dimensionful parameter $l_{ct}$ in the final analysis. 
    Whereas, in the strongly non-commutative regime, it is manifestly clear that there isn't any counterterm contribution at all.  
\end{itemize}

\begin{itemize}
    \item \textbf{Null-null junction contributions:} 
    The junction of the null boundaries contributes a codimension-2 term to the gravitational action, given by
    \begin{align}
        S_{\mathcal{X}}&= -\frac{1}{2\kappa_{10}^2}2\int d^8 X\, e^{-2(\phi-\phi_0)}\sqrt{\gamma}\ln\left|e^{-\frac{1}{2}(\phi-\phi_0)}\frac{n_+\cdot n_-}{2}\right|\Bigg|_{u=u_b}. \label{aniso-junct}
    \end{align}
    The metric on the junction cross-sections is
    \begin{align}
         d\gamma^2&= u^2(h(u)(dx^2+dy^2)+dz^2)+d\Omega_5^2.\nonumber
    \end{align}
 with the corresponding integration measure being $\sqrt{\gamma}=u^3h(u)\sqrt{g_{S^5}}$. Given that the null normal vectors have the components
 \begin{align}
      n_{\pm\mu}&=\left(-\alpha,  \frac{\mp\alpha}{u^2}, \vec{0}\right),\nonumber
 \end{align}   
  when plugged back into (\ref{aniso-junct}), yields
    \begin{align}
        S_{\mathcal{X}}=-\frac{2}{16\pi G_N}L^3u_b^3\ln\left|\frac{\alpha^2}{u_b^2(h(u_b))^{1/4}}\right|.\nonumber
    \end{align}
    Taking appropriate limits, the corresponding contribution in the near-AdS regime is 
    \begin{align}
        S_{\mathcal{X}}  &=\frac{L^3u_b^3}{16\pi G_N} \left(4\ln\frac{u_b}{\alpha}\right)+O(a^{4}), \label{aniso-junct}\\
        \end{align}
         And in strongly non-commutative regime is 
\begin{align}
        S_{\mathcal{X}}
        &=\frac{L^3u_b^3 }{16\pi G_N}\left(-2  \ln \left(\alpha a\right)\right)+O\left(a^{-4}\right).\label{aniso-NCJunct}
    \end{align}
\end{itemize}
\begin{itemize}
    \item In the weakly noncommutative regime:
\end{itemize}
Summing all perturbative contributions (\ref{aniso_AdSbulk}), (\ref{aniso-ct}), (\ref{aniso-junct}) in the regime where the UV cutoff dominates the non-commutative coupling, i.e., $a \ll 1/u_b$, reproduces the known result for the  total action complexity for pure AdS

\begin{align}
    \mathcal{C}_{\mathcal{A}}
    &=\frac{L^3}{16\pi^2 G_N\epsilon^3} \left( 4 \ln \left(\frac{3l_{ct}}{\alpha}\right)\right)+O(a^{4}).\label{aniso-AdSCA}
\end{align}
Where in the final expression, we have conventionally expressed UV cutoff $u_b$ in terms of  boundary lattice spacing $\epsilon$ via $u_b = 1/\epsilon$ .
\begin{itemize}
    \item In the strongly noncommutative regime:
\end{itemize}
 Conversely, in the opposite regime where the non-commutative coupling dominates, i.e., $a \gg 1/u_b$, the action complexity is given by the sum of \eqref{aiso-bulk}, \eqref{aiso-bound} and \eqref{aniso-junct}, that evaluates to be 
\begin{align}
    \mathcal{C}_{\mathcal{A}}
 &= { \frac{L^3 }{16\pi^2 G_N\epsilon^3}\left(-2   \ln (\alpha a)+\frac{20-4\alpha }{3}\right)+O\left({a^{-1}}\right).}\label{aniso-NCCA'}
\end{align}
The arguments of logarithm have been rendered dimensionless by rewriting them in  the units of AdS radius. 
It can be seen that for the anisotropic case of noncommutative field theory, the action complexity is positive only in the range $\ln (\alpha a)<\frac{10-2\alpha}{3}$.This feature of action complexity is parallels with the action complexity of the DDSYM theory. However, unlike the DDSYM theory, but instead just like the local SYM field theory, the UV divergence scales with volume $L^3$! We defer our speculations about this counterintuitive result till Sec. \ref{DO}.

\section{Noncommutative SYM theory: Isotropic case} \label{NCSYM-iso}
In the previous two cases of DDSYM and anisotropic NCSYM, the deformation had a preferred direction(s) and the full volume complexity turned out to be independent of the deformation parameter, i.e. turned out to be identical to undeformed SYM (pure AdS). We suspect this is a dimensional accident, i.e. artifact of the fact that there are only two  spatial directions along which the deformation is turned on, and if the the deformation democratically involves all spatial and temporal directions, the full volume complexity will exhibit features distinct from the undeformed theory. To confirm our conjecture, in this section, we consider the  Maldacena-Russo solution of a D3-brane in a constant NS $B$ field background given as \cite{Lin:2000ps},
\begin{align}    &ds_E^2=l^2\, {u^2}\,\Phi\,\left(\Phi^{-2}\left(dx_0^2+dx_1^2+dx_2^2+dx_3^2\right)+u^{-4}du^2+u^{-2}d\Omega^2_5\right)\nonumber\\
    \Phi&=(1+a^4 u^4)^{1/2}, \quad F_{0123u}=i4l^4u^3\Phi^{-4},\nonumber\\
    B_{01}=B&_{23}=\sqrt{g_s}a^2u^4\Phi^{-2}, \quad A_{01}=A_{23}=-i \frac{a^2l^2}{\sqrt{g_s}}u^4\Phi^{-2},\quad \chi=i\frac{a^4u^4}{g_s} \label{Mald-Russo sol}
\end{align}
In this case, the noncommutativity is involves all three spatial directions as is obvious from dual metric. One can perform the dimensional reduction of (\ref{Mald-Russo sol}) to read out the 5-dimensional background metrics in the Einstein frame as, 
\begin{align}
        ds^2 &=l^2u^{-2}\Phi^{8/3}du^2+l^2u^2\Phi^{2/3}(-dt^2+dx_1^2+dx_2^2+dx_3^2)
\end{align}
where $\Phi=(1+a^4u^4)^{1/2}$. We will work in spherical polar coordinates ($\rho,\theta,\phi$) instead of the cartesian $x_1,x_2,x_3$ and follow a similar procedure as above to compute the volume complexity.\\
\subsection{Volume complexity of the isotropic NCSYM}
The volume functional in the five-dimensional reduced bulk in the spherical polar coordinates mentioned above is,
\begin{align}
     V&=\int du\ d\rho \ d\theta \ d\phi \  l^4u^4 \Phi^{4/3}\rho^2 \sin \theta\sqrt{\frac{\Phi^2}{u^4}-t'^2(u)}=\frac{4\pi l^4 R^3}{3}\int_{0}^{1/\epsilon} du \ u^2\Phi^{7/3}
\end{align}
\begin{equation}
    \mathcal{C}_V=\frac{1}{3} \,\frac{l^3}{G^{(5)}_N} \,\frac{V_3}{\epsilon^3} \,\,_2F_1\left(-\frac{7}{3},\frac{3}{4};\frac{7}{4};-\left(\frac{a}{\epsilon}\right)^4\right) \label{CV anisoropic full}
\end{equation}
Here $V_3=4\pi R^3/3$ is the boundary spatial volume and $1/\epsilon$ is the bulk radial cutoff i.e. $a<\epsilon^{-1}$
\begin{equation}
\mathcal{C}_V=\frac{1}{3} \,\frac{l^3}{G^{(5)}_N} \,\frac{V_3}{\epsilon^3} \, \left[1+\left(\frac{a}{\epsilon}\right)^4+\frac{14}{33} \left(\frac{a}{\epsilon}\right)^8 +\ldots\right] \label{CV isoropic full}
\end{equation}
If we take the limit $a/\epsilon\to 0$ in (\ref{CV anisoropic full}), we can reproduce the pure AdS volume complexity obtained in \cite{Reynolds:2016rvl}.
Similarly, in the extreme non-commutative limit, i.e. $a/\epsilon\to \infty$, we get,
\begin{equation}
 \mathcal{C}_V=\frac{3}{37} \,\frac{l^3}{G^{(5)}_N} \,\frac{V_3}{\epsilon^3} {{\left(\frac{a}{\epsilon}\right)}}^{28/3} \left[ 1 +O\left(\frac{1}{(a/\epsilon)^4}\right)\right] \label{extreme-NC isotropic full}
\end{equation}
Here we note that the leading order of the volume complexity $\mathcal{C}_V\propto \left(\frac{1}{\epsilon}\right)^{37/3}\times R^{3}\times a^{28/3}$ which means there is a hyperscaling law w.r.t radial cutoff, $\frac{1}{\epsilon}$. Unlike the NCSYM and dipole case, the dependence of the non-commutative parameter $a$ further supports the argument that, the absence of $a$ in NCSYM and dipole cases is a dimensional accident.
\subsection{Action complexity (CA) for the isotropic case (5D)}
 For the case with self dual $B$ backgrounds, the 5-dimensional gravitational action can be derived by dimensionally reducing 10 dimensional IIB supergravity. The resulting 5 dimensional theory will have a dilaton profile $\Phi(u)$ and is shown to be the holographic dual to the NCYM with isotropic non-commutativity\cite{Lin:2000ps}. The bulk action of the dimensionally reduced 5-dimensional theory \footnote{$\frac{R^5\omega_5}{2\kappa^2_{10}}=\frac{1}{16\pi G_N^{(5)}}$} is  \begin{align}
   S_{gravity}    =  \frac{1}{16\pi G_N}\int d^5x\sqrt{-g}\left(\mathcal{R}-\frac{10}{3}\frac{(\nabla \Phi)^2}{\Phi^2}+\Phi^{-\frac{8}{3}}(20-8\Phi^{-4})\right),\nonumber
 \end{align}
 here, we have set AdS radius to be one, and the resulting gravitational background is found out to be the following
\begin{align}
    ds^2&=u^2(-dt^2+dx^2+dy^2+dz^2)\Phi^{2/3}+\frac{du^2}{u^2}\Phi^{8/3}, \hspace{1.5cm} \Phi(u)=(1+a^4u^4)^{1/2}.\nonumber
\end{align}
Ricci scalar can be shown to of the following form 
\begin{align}
      \mathcal{R}&=-\frac{20 \left(3 a^8 u^8+10 a^4 u^4+3\right)}{3 \left(a^4 u^4+1\right)^{10/3}}.\nonumber
\end{align}
Just like in the earlier sections, the WdW patch - the domain of integration of the bulk action is bounded by the null rays 
\begin{align*}
  dt^2 &=\frac{du^2}{u^4}\Phi^{2},\nonumber\\
   \int_{t_-}^{t_+}dt&=2\int^{u_b}_{u} du' \,\frac{\Phi(u')}{u'^2}.
\end{align*}
We have seen that the  gravitational action has many codimension higher terms coming from the various kind of boundaries, we start by evaluating the codimension-0 element which is same as the bulk contribution.
\begin{itemize}\item \textbf{Computation of bulk action: }
 The bulk action is evaluated over WdW patch with integration measure $\sqrt{-g}=u^3\Phi^{\frac{8}{3}}$ and gives
 
 \begin{align}
   S_{bulk}     &=  \frac{1}{16\pi G_N}\int d^5x\sqrt{-g}\left(\mathcal{R}-\frac{10}{3}\frac{(\nabla \Phi)^2}{\Phi^2}+\Phi^{-\frac{8}{3}}(20-8\Phi^{-4})\right),\nonumber\\  &=\frac{1}{16\pi G_N}\int d^3x\int_{0}^{u_b}du\int_{t_-}^{t_+} dt\,\, u^3\left(a^4 u^4+1\right)^{4/3}\left(\frac{8 \left(2-5 \left(a^4 u^4+1\right)^2\right)}{3  \left(a^4 u^4+1\right)^{10/3}}\right)\nonumber.
 \end{align}
 Working perturbatively, prior to capturing the non-commutative answer, we always first verify the AdS side of the story. Evaluating the bulk action in the perturbative limit when, the UV cutoff scale dominates the non-commuative coupling i.e. $a\ll 1/u_b$, reproduces the expected result for the known case of pure AdS 
  \begin{align}
      S_{bulk}&=  \frac{V_3u_b^3}{16\pi G_N}\left(-\frac{4}{3}+O(a^{4})\right).\label{aniso-AdSbulk}
        \end{align}
        Once, we have the correct match with the expected scaling of the known AdS result, we next evaluate the non-commutative answer in the perturbative limit when the non-commutative coupling dominates the UV cutoff scale or in other words, $a\gg 1/u_b$. We obtain,
        \begin{align}
               S_{bulk}&=\frac{V_3}{16\pi G_N}\left(-\frac{2}{3} a^2 u_b ^5 +O(a^{-2})\right).\label{iso-NCbulk}
        \end{align}
        We will comment on the scaling characteristics of the various contributions later when we will have put together all the pieces coming fr.m the various elements. Proceeding now over to the codimension-1 element, which is the contribution from the null boundaries of the WdW patch. 
        \end{itemize}
    \begin{itemize}
        \item \textbf{Null boundary components:} The boundary component contributing towards the gravitational action with null boundaries is of the sum over all the null boundary components of following form 
        \begin{align}
            S_{boundary }&=\sum_{j=\pm}\frac{2}{16\pi G_N}\int d^3x\int ds_{j}\,\sqrt{\gamma_{j}}\,\kappa_{j}.\label{iso-null}
        \end{align}
        Here,  the parameters $s$ on null generators are chosen to be increasing in the future direction,  $\gamma $ is the induced metric on the cross-sections of constant $s$. Moreover,  $n^{\mu}\nabla_{\mu} n^{\nu}=\kappa n^{\nu}$, where, $n$ is the null normal to the surface.  
Boundaries of WdW are given by 
 Null surfaces $\mathcal N_{\pm}$
   \begin{align}
       (t-T)&=\pm \int^{u_b}_{u} du' \,\frac{\Phi(u')}{u'^2}\nonumber.
   \end{align}
   In other words, null surfaces $\mathcal{N}_{\pm}$ are governed by 
   \begin{align}
   t&=T\pm \int^{u_b}_{u} du' \,\frac{\Phi(u')}{u'^2},\nonumber
\end{align}
where, $+/-$ labels the upper/lower null boundaries, respectively. The parameters on null generators are chosen to be increasing in the future direction and are taken to be $s_{\pm}=\mp u$. With this parametrization, the null normal vectors takes the following form 
\begin{align}
    n_{\pm}^{\mu}=\left(\frac{\alpha\Phi(u)}{ u^2},\mp\alpha,\vec{0}\right),\nonumber
    \end{align}
where $\alpha$ is some normalization constant, that cannot be fixed at this stage due to the null nature of the normals. This determines the quantities $\kappa $ through the following equation 
\begin{align}
  n_{\pm}^{\mu}\nabla_{\mu} n_{\pm}^{\nu}=\kappa_{\pm} n_{\pm}^{\nu},\nonumber
\end{align}
to be
 \begin{align}
    \kappa_{\pm}&=\mp\frac{10\alpha a^4  u^3}{3  \left(a^4 u^4+1\right)}\nonumber. 
\end{align}
Which, when summed over for both null boundary  components and plugged back in (\ref{iso-null}) governs 
\begin{align}
    S_{boundary }  &=\frac{1}{16\pi G_N}\frac{40\alpha V_3a^4}{3}\int_0^{u_b}du\,\frac{ u^6}{ \left(a^4 u^4+1\right)},\nonumber
\end{align}
where, on the sections of constant $s_{\pm}$, the induced metric is 
\begin{align}
d\gamma^2&= u^2(dx^2+dy^2+dz^2)\Phi^{2/3},\nonumber
\end{align}
giving the  integral measure of $\sqrt{\gamma}=u^3\Phi(u)$.
 Working perturbatively, we first verify the AdS side of the story by  evaluating the bulk action in the perturbative limit when, $a\ll 1/u_b$
    \begin{align}
      S_{boundary} &=  \frac{V_3}{16\pi G_N}a^4\alpha u_b^7\left(\frac{40}{21}+O(a^4)
      \right),
      \end{align}
      this correctly reproduces the expected result for the known case of pure AdS to the zeroth order in coupling $a$. Now looking ahead and working out the corresponding answer when the non-commutative coupling dominates the UV cutoff scale $a\gg 1/u_b$. We obtain,
      \begin{align}
         S_{boundary} &=\frac{V_3}{16\pi G_N}\left(\frac{40}{9}\alpha u_b^3+O\left(\frac{1}{a^4}\right)\right).\label{iso-NC bound}
    \end{align}
    We leave the analysis of the above for later when we had included all the action components. Also, normalization constant $\alpha$ will also be fixed in the final analysis.  
      \end{itemize}
\begin{itemize}
    \item \textbf{LMPS counter-term prescription}
It has been pointed out in \cite{Lehner:2016vdi} that the gravitational action depends upon the choice of the parametrization $s$ on the null boundary components. To restore the coordinate independence of the gravitational action, it is required to include the following counterterm component in the action, for each of the null boundaries.
\begin{align}
   \Delta S&=\sum_{j=\pm}\frac{-2}{16\pi G_N} \int d{A}_{j} \,ds_{j} \,{\Theta}_{j} \ln|l_{ct}{\Theta}_{j}| ,\hspace{3cm} \text{where,}\,\,{\Theta}_{\pm}\equiv \frac{1}{\sqrt{{\gamma}}}\frac{\partial \sqrt{{\gamma}}}{\partial s_{\pm}}.
\end{align}
Here, ${\gamma} $ is the induced metric on the constant $s$  sections. Since, metric is defined on the null section, it is a codim-2 surface.
And on the section of constant $s_{\pm}=\mp u$, the integration measure is the same as obtained earlier.\\
It can be seen that, $ \Theta_+=-\Theta_-.$ Therefore, we just label  $ \Theta_-$ by  $ \Theta$. Hence,
\begin{align}
   \text{ }\,\, \Theta
&=\frac{5 a^4 u^4+3}{ u \left(a^4 u^4+1\right)}.\nonumber
\end{align}
It turns out that after accounting for various signs coming from the causal structure of the null components, the contributions for  both the surfaces are the same.  Therefore, the total contribution from the counter term is twice the contribution of the other one. Hence, their sum is
\begin{align}
     \Delta S &=\frac{4V_3}{16\pi G_N}\,\int^{\infty}_{0} du\,\sqrt{\gamma}\,{\Theta} \ln|l_{ct}{\Theta}|\nonumber,\\
    &=\frac{4V_3}{16\pi G_N}\,\int_0^{u_b}du \,u^2\frac{5 a^4 u^4+3}{ \left(a^4 u^4+1\right)^{1/2}}\ln\left|\frac{l_{ct}(5 a^4 u^4+3)}{ u \left(a^4 u^4+1\right)}\right|\nonumber.
\end{align}
We have included the dimensionful factor of $l_{ct}$ to make the argument of logarithm dimensionless. This $l_{ct}$ will be deduced later from the consistency with the known AdS result in terms of normalization constant $\alpha $ in the final analysis.  
In the appropriate perturbative regimes we obtain the following contribution about  AdS regime 
\begin{align}
  \Delta S &=\frac{V_3u_b^3}{16\pi G_N}\left[\frac{4}{3}+4 \ln \left(\frac{3l_{ct}}{ u_b}\right)\right]+O(a^{4}).\nonumber\end{align}
  and in the strongly non-commutative regime to be 
  \begin{align}
       \Delta S
    &=\frac{V_3 
    }{16\pi G_N}\left(2 a^2 u_b ^5 \left(- \ln \left(\frac{u_b}{l_{ct}} \right)+ \ln \left(5\right)+1\right)\right)+O(a^{-2}).\label{iso-NC count}
\end{align}
 \end{itemize}
\begin{itemize}
    \item \textbf{Null-null junction } The null boundaries meet along the null junction, which also gives rise to the codimension-2 components contributing to the action complexity of the following form 
    \begin{align}
         S_{\mathcal{X}}&= \frac{-2}{16\pi G_N}\int d^3 x\, \sqrt{\gamma}\ln\left|\frac{n_+\cdot n_-}{2}\right|\Bigg|_{u=u_b}\nonumber.
    \end{align}
    The metric on the sections of the junctions is
\begin{align}
d\gamma^2&= u^2(dx^2+dy^2+dz^2)\Phi^{2/3}\nonumber, 
\end{align}
With the integral measure $\sqrt{\gamma}=u^3\Phi(u)$.
Given that the various normals have the following  components, 
\begin{align}
      n^{\mu}_{\pm}&= \left(\frac{\alpha \Phi}{u^2}, \mp\alpha , \Vec{0}\right)\nonumber, 
\end{align}
Plugging this all into $S_{\mathcal{X}}$ we get the following 
\begin{align}
   S_{\mathcal{X}}&=  \frac{-2}{16\pi G_N}\int d^3 X\, \sqrt{\gamma}\ln\left|\frac{n_+\cdot n_-}{2}\right|\Bigg|_{u=u_b}\nonumber,\\
&= \frac{-2V_3u_b^3(1+a^4u_b^4)^{1/2}}{16\pi G_N}\ln\left(\frac{\alpha ^2 \left(a^4 u_b^4+1\right){}^{8/3}}{u_b^2}\right)\nonumber.
\end{align}
Simply taking appropriate limits, renders the following contributions, first in the near AdS regime we get 
\begin{align}
      S_{\mathcal{X}}   &=\frac{V_3u_b^3}{16\pi G_N}\left(4 \ln \left(\frac{u_b}{\alpha}\right)+O(a^4)\right).
      \end{align}
      where as in the strongly non commutative regime to be 
      \begin{align}
      S_{\mathcal{X}}
    &=\frac{V_3  u_b^3}{16\pi G_N}\left(-\frac{64}{3} a^2 u_b^2 \ln \left(a u_b\right)+4 a^2 u_b^2 \ln \left(\frac{  u_b}{\alpha}\right)\right)+O(a^{-2}).\label{iso-NC junct}
\end{align} 
\end{itemize}
\begin{itemize}
    \item In the weakly noncommutative regime: 
\end{itemize}
Summing over all the perturbative contributions obtained when  the UV cutoff scale dominates the non-commuative coupling i.e. $a\ll 1/u_b$, consistently reproduces the total action complexity for the known case of pure AdS to be

    \begin{align}
          \mathcal{C}_{\mathcal{A}}&=\frac{V_3}{16\pi^2 G_N\epsilon^3}\left(4\ln\left(\frac{3l_{ct}}{\alpha}\right)\right)+O(a^4).
    \end{align}
    Where in the end, we have expressed $u_b=\frac{1}{\epsilon}.$
    \begin{itemize}
        \item In the extreme noncommutative regime: 
    \end{itemize}
   On the other hand,  in the extreme noncommutative regime i.e. $a\gg 1/u_b$ the action complexity is given by the sum of \eqref{iso-NCbulk}, \eqref{iso-NC bound}, \eqref{iso-NC count} and \eqref{iso-NC junct} to be 
 \begin{align}
    \mathcal{C}_{\mathcal{A}}    &=\frac{a^2V_3 }{16\pi^2 G_N\epsilon^5} \left(-\frac{64}{3}  \ln \left(a \right)+\frac{58}{3}  \ln \left(\epsilon\right)+2 \ln \left( 5\right)+\frac{4}{3}+\ln(l_{ct})-4\ln (\alpha)\right)+O(1). 
 \end{align}
 Here we, as we did in the previous sections, in the end we have rescaled every argument of the logarithm in the units of AdS radius.  It is evident that the overall action complexity for the noncommutative case is negative unless the noncommutativity parameter satisfies the bound $\ln(\frac{a^{64}\alpha^{12}}{5^6l_{ct}^3\epsilon^{58}}) <4$. Also,  the leading order scaling with the noncommutative coupling is the product of quadratic and logarithmic scaling with the nonlocal coupling parameter. It is however clear that due to the involvement of the UV scale the complexity cannot be made positive in any range of values in the coupling space. This is hinting at  the theory being  UV incomplete, i.e. it has UV instability (in past work \cite{Katoch:2023dfh} we have also noted that negative holographic complexity could imply a system without a stable ground state).   Also, the action complexity in the noncommutative case scales with the fifth power of the UV cutoff, unlike the local field theory  that is expected to scale cubically in the UV cutoff scale. This power law departure from the extensive scaling is interpreted as the effect of nonlocality on the action complexity. 
\section{Discussions and Outlook} \label{DO}

Our work was motivated by past work \cite{Barbon:2008ut, Karczmarek:2013xxa, Shiba:2013jja, Pang:2014tpa} displaying universal volume  divergence scaling laws for the Holographic Entanglement Entropy for nonlocal field theories admitting holographic duals as opposed to area laws for local field theories. In this work we have considered the following three case: a dipole deformed SYM theory (DDSYM), and the anisotropic and isotropic noncommutative SYM theories (NCSYM). All these are obtained by nonlocal irrelevant deformations of the SYM theory and admit holographic gravity duals: 10d SUGRA backgrounds with nontrivial dilaton and NS-NS B-fields turned on. (However there is a crucial distinction between the dipole-deformation and noncommutative deformation. In the case of the DDSYM theory, there is no UV-IR mixing while for the NCSYM there is the well-known phenomenon of UV-IR mixing).  In our present work we established that the holographic (volume and action) complexity does exhibit violations of volume law scaling in the UV divergence, but  does not exhibit \emph{universality} in the UV divergence scalings. For the dipole deformed SYM theory (DDSYM) and the anisotropic noncommutative SYM theory (anisotropic NCSYM), the UV divergences in the volume scale cubically as opposed to linearly with the extension along the dipole deformation (linear scaling being equivalent to a volume law). For the isotropic NCSYM theory the UV divergence scales with an exponent $37/3$ for volume complexity while it scales with exponent $5$ in the action complexity. So even for the isoptropic case, the ``hypervolume"  divergence law is not universal. These results are in sync with and extend our past work on lower dimensional UV-deformed nonlocal CFTs such as LST and WCFT \cite{Chakraborty:2020fpt, Katoch:2022hdf, Bhattacharyya:2022ren}.\\

 Now we discuss the subregion volume complexity results. For the DDSYM case, we clearly observe a phase transition, mirroring the behavior of holographic EE (as well as other probes such as Wilson loops and low order correlation functions as have been shown in the past literature). The critical subregion size for both subregion complexity and entanglement entropy (worked out in the appendix) turns out to be the same thereby establishing a unanimous phase transition scale (length) for the onset of nonlocality (dipole interaction dominated physics). For subregions larger than the critical subregion, the subregion volume complexity scales linearly with subregion extension along the dipole deformation (same as what is expected for a local SYM theory), however for subregions smaller than the critical size (length), the subregion volume complexity scales cubically with the subregion extension along the dipole deformation.  For the anisotropic NCSYM theory, we also observe a phase transformation with similar traits. There are two phases: for subregions larger than the critical subregion, the subregion volume complexity scales linearly with subregion extension along the noncommutative deformation (same as what is expected for a local SYM theory), however for subregions smaller than the critical size (length), the subregion volume complexity scales cubically with the subregion extension along the noncommutative deformation. The phase transition point, i.e. the critical subregion length can be extracted from the plots which agrees within numerical errors with the analytical expression obtained from the entanglement entropy transition point. These results are very similar the LST counterparts we studied previously \cite{Chakraborty:2020fpt, Katoch:2022hdf} where the subregion volume complexity exhibited the Hagedorn-like phase transition set the scale of the deformation parameter(s). For the isotropic NCSYM we were unable to get numerical plots to work out reliably and conclusively as there were runaway errors even for very small values of the noncommutative deformation parameter $a$ (these are discussed in details in appendix \ref{app: num_error}). This is true not just for the subregion volume complexity but also for the holographic entanglement entropy. We plan to revisit this issue in the future.\\

Next we discuss the action complexity results. The action complexity results in the weakly nonlocal (near AdS) limit are in sync with the volume complexity, i.e. leading UV divergence piece match (upto numerical factors involving spacetime dimensions). The subleading pieces, however are not in one-to-one correspondence with the subleading divergences in volume complexity. This is not a surprise and has been documented in the literature since the initial days of investigations into holographic complexity.  Now one generic feature of the action complexity results, inferred from by inspecting the far AdS/ highly nonlocal limit of the action complexity. For the action complexity to remain positive, the nonlocal deformation parameter(s) must have an upper bound. This is reminiscent of our past investigations of Little String theories (LST) \cite{Katoch:2022hdf} where the nonlocal irrelevant deformation couplings had to satisfy bounds for the complexity to remain real positive. There the bound reflected the fact that the LST with the irrelevant deformations was not UV complete. We believe a similar scenario at play here: the nonlocal theories we consider might not be UV complete for arbitrary large values of the nonlocal deformation parameter. For the DDSYM this upper bound puts a restriction on the 't-Hooft coupling! For the NCSYM theory, both the anisotropic and isotropic case, examining the action complexity in the extreme noncommutative regime and demanding the complexity remain positive leads to an upper bound on the noncommutativity parameter (to be precise, the dimensionless ratio of the noncommutativity parameter and AdS radius). However, there is a clear distinction between the anisotropic and isotropic case: for the anisotropic case the upper bound is independent of the UV cutoff, while for the isotropic case the bound is a function of the UV cutoff! In the other limit, i.e. weakly noncommutative regime, the action-complexity leads to perturbative corrections to the local SYM theory (pure AdS) action complexity. Although physically not that interesting, this furnishes a nice consistency check of our action complexity results (they all reproduce the local SYM theory results in the weak deformation regime). We also want to point out to a technical subtlety regarding the action complexity computations. For the action complexity computations, we have been using a 10d type IIB SUGRA action and in such case one has to incorporate a topological term \cite{Kurlyand:2022vzv,Apolo:2025wcl}. Otherwise the action complexity does not reproduce the expected answer in the pure AdS limit. If instead one uses a dimensionally reduced effective 5d SUGRA action, then one does not need to incorporate this term.\\

Next we discuss some limitations of or exceptions to our particular approach and the broader framework. In the entirety of our work we have assumed that the leading UV divergence structure, in particular the scaling exponents, of complexity are unambiguous indicators of whether the underlying theory has local or nonlocal dynamics: volume extensive scaling indicating locality while hypervolume scaling indicating nonlocality. And indeed in the past literature, this approach has been utilized for holographic probes such as sugra correlators, Wilson loops, Entanglement entropy (extremal surfaces), this appears to the case, the scaling exponents change as one moves from a local theory in the IR to a nonlocal theory in the UV. However, when working with a probe such as complexity, one has to be extra cautious and bear in mind that are a multitude of different prescriptions for defining quantum complexity given a Hilbert space of a quantum system, and there is no guarantee the leading UV divergence will be identical for all of them. So while working complexity, it is entirely plausible that leading UV divergences are also distinguishing different prescriptions
for complexity (both in the field-theory complexity), rather than distinguishing actual physical phases governed by local/nonlocal physics. This feature is unique to complexity, as other probes such as CFT correlators, or Wilson loops or entanglement entropy are defined unambiguously in the boundary quantum theory. Another, related issue in holographic setups is regarding the identification or interpretation of local dynamics in the gravity (bulk) with nonlocal quantum computation in the dual field theory \cite{May:2019yxi} using shared entanglement. As such, the complexity of computational tasks in the boundary involving nonlocal quantum computation might not scale extensively in the boundary at all, refer to \cite{Bluhm:2025qgd, Allerstorfer:2023ycc, Asadi:2024fda} for some work on the complexity theory for nonlocal quantum computations in quantum mechanics and field theory in the context of holography.\\

Finally, we discuss few avenues for future work. The most natural and immediate extension would be to include the effects of finite temperature and finite charge/chemical potential in complexity for nonlocal field theories using the holographic charged black hole/brane SUGRA backgrounds instead \cite{Eccles:2021zum, Couch:2017yil}. This is always an interesting exercise to carry out because the dual theory is nonlocal and it has the potential to shed light on novel physical effects associated purely with finite temperatures which might not appear in the local theories or for which we might not have any intuition from local theories. A second avenue to pursue is the subregion complexity for the isotropic NCSYM case for which the numerical estimates obtained using Mathematica or Python were unstable and unreliable for even small or moderate values of the deformation parameter $a$ - the numerical errors rapidly escalate to large values in course of the numerical solution of the RT curve. The numerics do not reproduce the pure AdS quadratic scaling of holographic entanglement entropy itself, let alone subregion complexity.  It would be desirable to come up with a numerical scheme where the errors could be brought under control for moderate to large values of the deformation parameter $a$ and tie up this loose end in our present work. Another instructive, although perhaps less interesting avenue to explore will be to compute the subregion action complexity for all the cases considered in this paper. Finally, casting a wider net, one can investigate the holographic complexity (in addition to entanglement entropy) properties of more general nonlocal field theories dual to string backgrounds which are not asymptotically AdS, e.g. the nonlocal (C)FTs dual to null boundary holography \cite{Ferko:2025elh,Dei:2025ryd,Chakraborty:2025rfm}. These are prototypes of celestial holography, and will help us explore holography/quantum gravity in backgrounds beyond the asymptotically AdS paradigm.

\section*{Acknowledgements} 
We thank the organizers and participants of the BIRS-CMI workshop on  \href{https://www.birs.ca/events/2025/5-day-workshops/25w5386}{``Quantum Gravity and Information Theory: Modern Developments"}  (9-14 Nov. 2025) which offered us a chance to present our results. We also thank Juan F. Pedraza for drawing our attention to the relevant works \cite{Fischler:2013gsa, Couch:2017yil, Eccles:2021zum}. Finally, we thank the anonymous referee for several insightful comments which improved our manuscript appreciably and for drawing our attention to the highly interesting works \cite{Allerstorfer:2023ycc, Asadi:2024fda, Bluhm:2025qgd}. The work of SP and SR is supported by the CRG grant of the Anusandhan National Research Foundation (ANRF), Department of Science and Technology (DST), CRG/2023/001120 (``Many facets of complexity: From chaos to thermalization"). The work of SR is also supported by the IIT Hyderabad seed grant SG/IITH/F171/2016-17/SG-47 as well the IITH funds RDF/IITH/F171/SR. The work of GK is supported by IIT Indore Sponsored Research Project no. CRG/2023/000904 (``From Modular Hamiltonian to Operator Algebra in Gauge/Gravity Dictionary").

\appendix
\section{Entanglement entropy results for dipole deformed SYM}\label{EE_dipole}
As the integral involved in area \eqref{eq: DT_area} is analytically not tractable, and we proceed numerically and study the behavior of area with subregion length \eqref{eq : DT_subregionlength} in figure \ref{fig:areavslength}. This shows a phase transition in entanglement entropy. This transition has been already observed in \cite{Karczmarek:2013xxa} for a near boundary (large $u$) RT surface. Here we have done a full numerical analysis and observed that as subregion length increase the entanglement entropy (area) becomes independent of subregion length as well as the deformation parameter. In the near AdS region when $a u<<1$, and the leading term in entanglement entropy is same as that of pure AdS. For small sub-region length degree of freedom inside the strip will be entangled to the degree of freedom outside the strip and the entanglement entropy will be proportional to the subregion length $l$. A limiting analysis of the area agrees with this interpretation.\\
\begin{figure}
    \centering
    \includegraphics[width=0.5\linewidth]{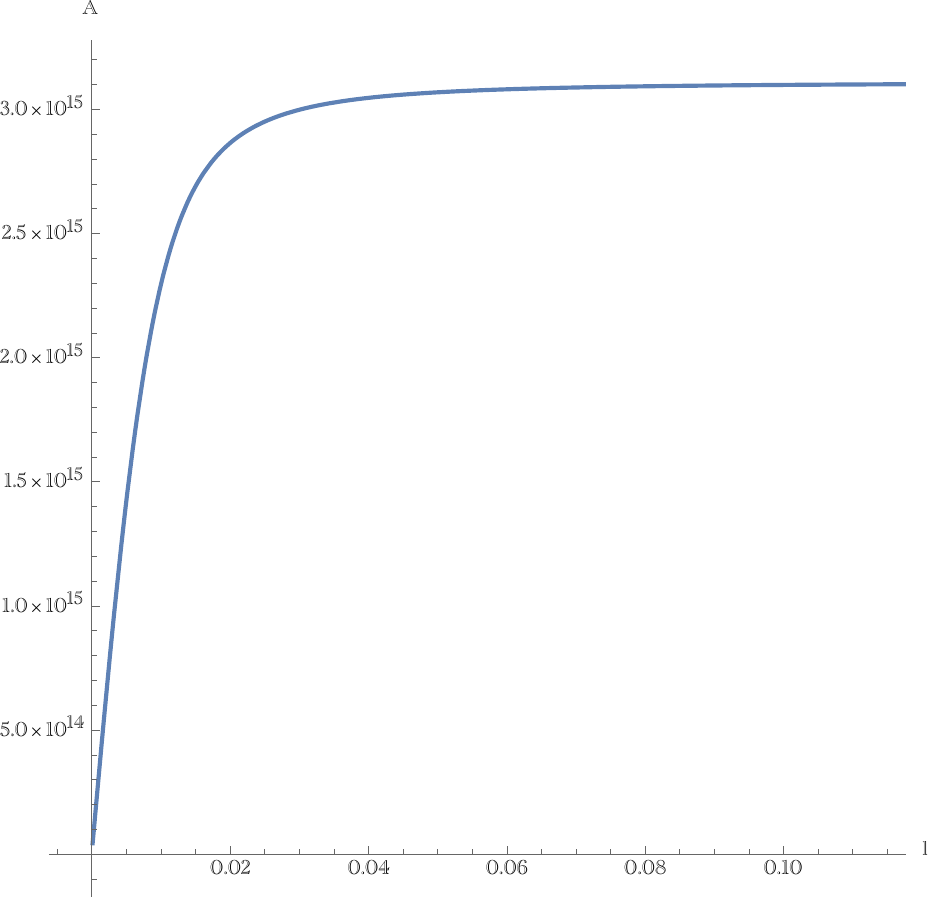}
    \caption{Plot of area vs subregion length for dipole deformed theory for $a=10^{-3.0}$ and $u_{b}=10^{2}$}
    \label{fig:areavslength}
\end{figure}
\newline
\textbf{For near AdS region (weak nonlocality) $(au<<1)$:}
\begin{eqnarray}
    \mathcal{A}=\frac{2 \pi ^{7/2} u_*^2 W^2 \Gamma \left(\frac{11}{6}\right)}{5 \Gamma \left(\frac{4}{3}\right)}-\frac{2 \pi ^3 W^2 \sqrt{\frac{1}{\epsilon ^6}-u_*^6} \, _2F_1\left(1,\frac{4}{3};\frac{11}{6};\frac{1}{\epsilon ^6 u_*^6}\right)}{5 u_*^6 \epsilon ^5}.
\end{eqnarray}
For $\epsilon\rightarrow0$ limit,
\begin{eqnarray}
    \mathcal{A}=\frac{\pi ^3 W^2}{\epsilon ^2}+\frac{4 \pi ^{9/2} W^2 \Gamma \left(\frac{2}{3}\right)^2 \Gamma \left(\frac{5}{6}\right)}{l^2 \Gamma \left(\frac{1}{6}\right)^2 \Gamma \left(\frac{1}{3}\right)}+\frac{8 (-1)^{2/3} \pi ^{9/2} W^2 \Gamma \left(\frac{2}{3}\right)^2 \Gamma \left(\frac{5}{6}\right)}{\sqrt{3} l^2 \Gamma \left(\frac{1}{6}\right)^2 \Gamma \left(\frac{1}{3}\right)}
\end{eqnarray}
\newline
\textbf{Far from AdS region (strong nonlocality) $(au>>1)$:}
The entanglement entropy is given by,
\begin{eqnarray}
   \mathcal{A}= \frac{2 \pi ^3 a W^2 }{3 \epsilon ^3}\sin \left(\frac{3 l}{2 a}\right)
\end{eqnarray}
For small sub-region $\frac{l}{a}<<1$
\begin{eqnarray}
   \mathcal{A}= \frac{ \pi ^3  W^2 l}{ \epsilon ^3} 
\end{eqnarray}
\section{Numerical errors in Isotropic NCSYM}\label{app: num_error}
If we proceed to study subregion complexity in the isotropic NCSYM case, the first step is to find the RT surface, which we take $\rho=\rho(z)$, so we will have the area as,
\begin{align}
    A=4\pi l^3\int_{0}^{z_0}\frac{dz}{z^3}\Phi \rho^2\sqrt{\Phi^2+\rho'^2}
\end{align}
After performing variational analysis over this area functional we have the second order DE with the boundary conditions as $\rho(z_0)=0$, and $\rho'(z_0)=\infty$ as,
\begin{align}
    \rho''-\frac{2\left(1+\frac{a^4}{z^4}+\rho'^2\right)}{\rho}-\frac{\left(5a^4z^3+3z^7\right)\rho'^3}{(a^4+z^4)^2}-\frac{3\rho'}{z}=0
\end{align}
Solving this equation numerically, with the boundary conditions we found that the solution is not compatible with the pure AdS limit.
To resolve this issue we focus on the AdS case ($a\rightarrow0$) for which we have the DE as,
\begin{align}
    \rho''-\frac{2(1+\rho'^2)}{\rho}-\frac{3\rho'^3}{z}-\frac{3\rho'}{z}=0.
\end{align}
It is known that the solution for this DE is $\rho^2+z^2=b^2$ \cite{Ryu:2006bv}.

Upon numerically solving this equation, we find that the solution takes the form of a semicircle only within a certain range of the parameter $z_{0}$. Outside this range, the solution deviates from a semicircular profile, as illustrated  in Fig.~\ref{fig:solutions_isotropic}.
\begin{figure}[htbp]
\centering
\begin{minipage}{0.32\textwidth}
    \centering
    \includegraphics[width=\linewidth]{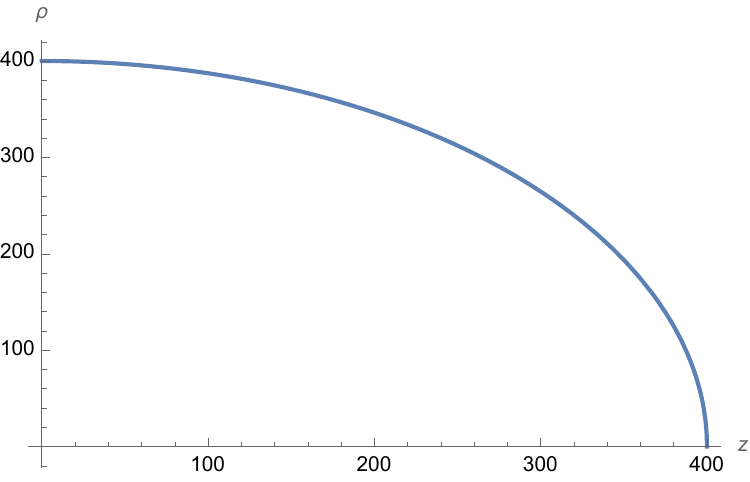}
\end{minipage}\hfill
\begin{minipage}{0.32\textwidth}
    \centering
    \includegraphics[width=\linewidth]{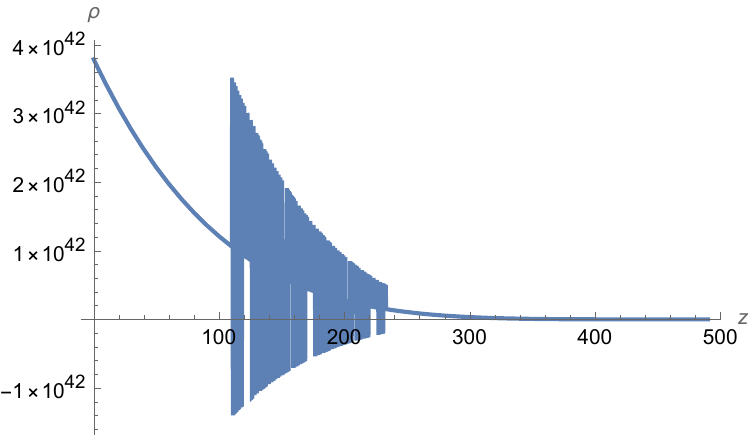}
\end{minipage}\hfill
\begin{minipage}{0.32\textwidth}
    \centering
    \includegraphics[width=\linewidth]{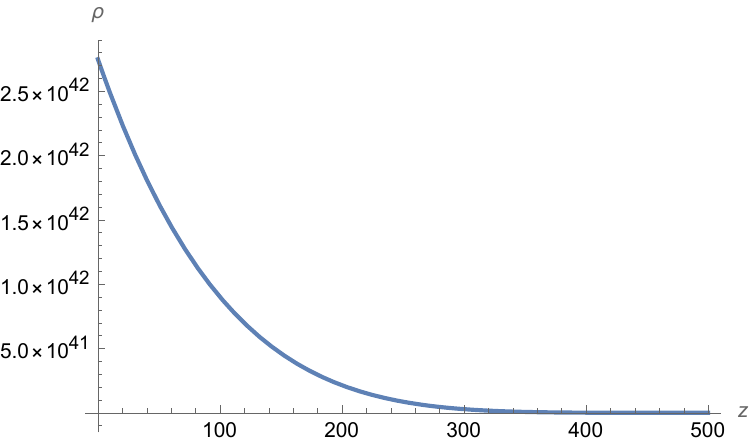}
\end{minipage}
\caption{The extremal surface with extremal points $z_{0}=400,490,500$ respectively}
\label{fig:solutions_isotropic}
\end{figure}
 This deviation is also reflected in the slope of the numerical plot of $\ln(A)$ versus $\ln(l)$. While the expected slope is 2, as the area $A\sim l^{2}$ for the pure AdS case. Our numerical results show a clear departure from this value, as demonstrated in the figure below.
\begin{figure}
    \centering
    \includegraphics[width=0.5\linewidth]{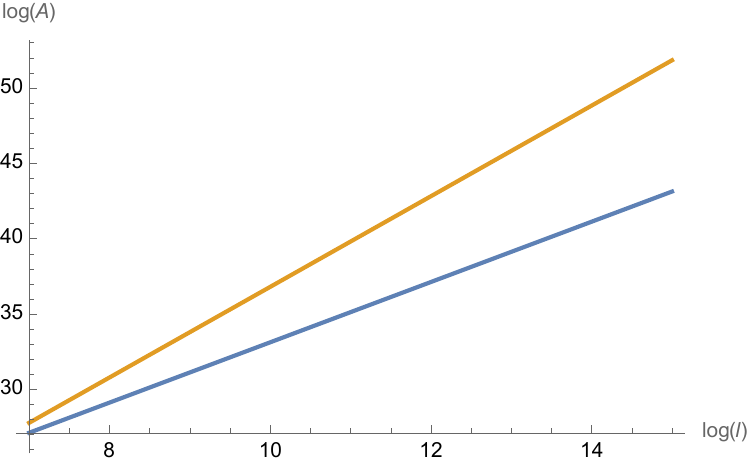}
    \caption{The blue curve which has been plotted for the range of $z_{0}$ over which we get semi-circle solution we have slope two. For the orange curve which is for a  broader range of $z_{0}$ we get slope value 3.}
    \label{fig:iso_pureads_issue}
\end{figure}
\newpage

\bibliography{references}
\bibliographystyle{JHEP}
\end{document}